\documentclass[
10pt,
showpacs,preprintnumbers,nofootinbib,
 amsmath,amssymb,
 aps,
prc,twocolumn,groupedaddress,superscriptaddress,
showkeys
]{revtex4-1}

\usepackage{graphicx}
\usepackage{dcolumn}
\usepackage{bm}
\usepackage[colorlinks=true,urlcolor=blue,citecolor=blue,breaklinks=true]{hyperref}

\newcommand{\leftsub}[2]{\vphantom{#2}_{#1}{#2}}

\newcommand{\cut}[1]{}

\newcommand{\ket}[1]{\ensuremath{| #1 \rangle}}
\newcommand{\bra}[1]{\ensuremath{\langle #1  |}}

\newcommand{\elem}[2]{\ensuremath{{}^{#2}\text{#1}}}

\newcommand{\clebsch}[6]{ \ensuremath{\left(\!\!
\begin{array}{cc}
 {#1} & \!\!\!\!{#2} \\
 {#4} & \!\!\!\!{#5} 
\end{array}
 \!\!\right|
\left.\!\!\!
\begin{array}{c}
 {#3}\\
 {#6}
\end{array}\!\! \right)}
}

\begin{document}


\title{{\em Ab initio} many-body calculations of nucleon-$^4$He scattering with three-nucleon forces}

\author{Guillaume Hupin}
 \email{hupin1@llnl.gov}
 \affiliation{Lawrence Livermore National Laboratory, P.O. Box 808, L-414, Livermore, California 94551, USA}
\author{Joachim Langhammer}%
 \email{joachim.langhammer@physik.tu-darmstadt.de}
 \affiliation{Institut f\"ur Kernphysik, Technische Universit\"at Darmstadt, D-64289 Darmstadt, Germany}%
\author{Petr Navr\'atil}
 \email{navratil@triumf.ca}
 \affiliation{TRIUMF, 4004 Wesbrook Mall, Vancouver, British Columbia, V6T 2A3, Canada}
\author{Sofia Quaglioni}
 \email{quaglioni1@llnl.gov}
 \affiliation{Lawrence Livermore National Laboratory, P.O. Box 808, L-414, Livermore, California 94551, USA}
\author{Angelo Calci}
 \email{angelo.calci@physik.tu-darmstadt.de}
 \affiliation{Institut f\"ur Kernphysik, Technische Universit\"at Darmstadt, D-64289 Darmstadt, Germany}
\author{Robert Roth}
 \email{robert.roth@physik.tu-darmstadt.de}
 \affiliation{Institut f\"ur Kernphysik, Technische Universit\"at Darmstadt, D-64289 Darmstadt, Germany}

\date{\today}

\begin{abstract}
We extend the {\it ab initio} no-core shell model/resonating-group method to include three-nucleon ($3N$) interactions for the description of nucleon-nucleus collisions. We outline the formalism, give algebraic expressions for the $3N$-force integration kernels, and discuss computational aspects of two alternative implementations. The extended theoretical framework is then applied to nucleon-$^4$He scattering using similarity-renormalization-group (SRG) evolved nucleon-nucleon plus three-nucleon potentials derived from chiral effective field theory. We analyze the convergence properties of the calculated phase shifts and explore their dependence upon the SRG evolution parameter.  We include up to six excited states of the \elem{He}{4} target and find significant effects from the inclusion of the chiral $3N$ force, e.g., it enhances the spin-orbit splitting between the $3/2^-$and $1/2^-$ resonances and leads to an improved agreement with the phase shifts obtained from an accurate $R$-matrix analysis of the five-nucleon experimental data. We find remarkably good agreement with measured differential cross sections at various energies, while analyzing powers manifest larger deviations from experiment for certain energies and angles.
\end{abstract}

\pacs{21.60.De, 25.10.+s, 27.10.+h, 27.20.+n}
\maketitle


\section{\label{sec:introduction} Introduction}

Recent progress in {\em ab initio} nuclear theory has been helping us reach a basic understanding of nuclear properties while paving the way to accurate predictions in the domain of light nuclei. This has been made possible by simultaneous advances in the fundamental description of the nuclear interaction, many-body techniques, and scientific computing. Today, accurate nucleon-nucleon ($NN$) and three-nucleon ($3N$) interactions from chiral effective field theory ($\chi$EFT)~\cite{Epelbaum2009,Machleidt2011} offer a much desired link to the underlying theory of quantum chromodynamics at low energies. At the same time,  a first-principles solution of the many-body problem starting from realistic Hamiltonians is not only being achieved for well-bound states~\cite{Pieper2001,Pastore2013,Navratil2007a,Hagen2012,Epelbaum2011}, but also is becoming possible for scattering and reactions as successful {\em ab initio} bound-state techniques are being extended to the description of dynamical processes between light nuclei~\cite{Nollett2007,Quaglioni2008,Hagen2012a,Baroni2013}. In techniques based on large-scale expansions over many-body basis states, this success is in part enabled by the use of similarity renormalization group (SRG)~\cite{Gazek1993,Wegner1994, Bogner2007, Hergert2007} transformations of the input Hamiltonian, where interactions can be softened in exchange for induced many-body terms~\cite{Jurgenson2009,Roth2011,Jurgenson2013, Roth2010}.   

One of the emerging techniques in the area of {\em ab initio} light-nucleus reactions is the no-core shell model combined with the resonating-group method or NCSM/RGM~\cite{Quaglioni2008,Quaglioni2009}. Here, RGM~\cite{Wildermuth1977,Tang1978,Fliessbach1982,Langanke1986,Lovas1998,Hofmann2008} expansions in $(A{-}a,a)$ binary cluster wave functions, where each cluster of nucleons is described within the {\em ab initio} NCSM~\cite{Navratil2000a,Navratil2000, Navratil2009, Barrett2013}, are used to describe the dynamics between nuclei made of interacting nucleons starting from realistic Hamiltonians. In the recent past, this technique has been successfully applied to compute nucleon~\cite{Navratil2010} and deuteron~\cite{Navratil2011} scattering on light nuclei, based on accurate $NN$ potentials obtained by SRG softening of the $\chi$EFT $NN$ potential at next-to-next-to-next-to-leading order (N$^3$LO) by Entem and Machleidt~\cite{Entem2003}. In these first applications, the omission of many-body forces induced by the renormalization of the input $NN$ potential introduced a dependence on the SRG resolution scale $\lambda$. Also neglected was the $3N$ component of the initial chiral Hamiltonian.    
Nevertheless, by choosing an appropriate value of $\lambda$ that reproduced the observed particle separation energies, the NCSM/RGM was capable of providing a promising realistic description of scattering data and even complex reactions such as the $^7$Be$(p,\gamma)^8$B radiative capture~\cite{Navratil2011a} or the $^3$H$(d,n)^4$He and $^3$He$(d,p)^4$He fusion rates~\cite{Navratil2012}. In addition, nucleon-nucleus NCSM/RGM wave functions combined with NCSM eigenstates of the composite $A$-nucleon system have been successfully used to compute the low-lying spectrum of the unbound $^7$He nucleus within the more complete framework of the no-core shell model with continuum (NCSMC)~\cite{Baroni2013,Baroni2013a}. However, a truly accurate {\em ab initio} description demands the inclusion of both induced and initial chiral $3N$ interactions.

In this paper, we present an extension of the  {\em ab initio} NCSM/RGM to include explicit $3N$-force components of the Hamiltonian in the description of nucleon-nucleus collisions, and discuss two alternative implementations of the approach. The extended formalism is then applied to the study of nucleon-$^4$He scattering using SRG-evolved $NN+3N$ Hamiltonians derived from the N$^3$LO $NN$ interaction of Ref.~\cite{Entem2003} along with the local form of the chiral $3N$ force at next-to-next-to-leading order (N$^2$LO) of Ref.~\cite{Navratil2007} entirely constrained in the two- and three-nucleon systems~\cite{Gazit2009}. We account for target-polarization effects by including, besides the ground state, up to six excited states of the $^4$He nucleus. We present a systematic investigation of the convergence of the computed phase shifts with respect to the parameters of the NCSM/RGM model-space, and explore their dependence upon the SRG resolution scale $\lambda$. 
The calculated phase shifts and derived angular cross sections and analyzing powers are compared to selected five-nucleon experimental data. We show the effect of the $3N$ force induced by the SRG technique and observe that the initial chiral $3N$ interaction enhances the spin-orbit splitting of the $^2P_{3/2}$ and $^2P_{1/2}$ resonances, leading to an improved agreement with the phase shifts obtained from an accurate $R$-matrix analysis of the five-nucleon experimental data~\cite{Hale} and angular distributions at intermediate nucleon incident energies.

An {\em ab initio} investigation of the elastic scattering of neutrons on $^4$He starting from realistic $NN+3N$ Hamiltonians has been obtained in the past within the Green's Function Monte Carlo (GFMC) method~\cite{Nollett2007}. The present study extends also to proton-$^4$He scattering, covers a wider range of energies,  and is the first to be performed using two- plus three-body Hamiltonians from $\chi$EFT. In particular, aside from a study of the photon-induced disintegration of the $^4$He nucleus~\cite{Quaglioni2007}, the present constitutes one of the first applications of chiral $NN+3N$ interactions to the {\em ab initio} calculation of scattering observables in systems with more than three nucleons together with the recent work of Refs.~\cite{Viviani2013} and~\cite{Viviani2013a}, and the first one for $A>4$. 

The paper is organized as follows.
In Sec.~\ref{sec:Formalism}, we provide a brief overview of the NCSM/RGM formalism and present algebraic expressions for the contribution of the $3N$ interaction to the Hamiltonian kernel. The application of this formalism to the elastic scattering of nucleons on $^4$He is presented in Sec.~\ref{sec:applications}. Finally, conclusions and outlook are given in Sec.~\ref{sec:Conclusions}.


\section{\label{sec:Formalism}{\em Ab initio} NCSM/RGM}

The general theoretical framework of the {\em ab initio} NCSM/RGM approach was introduced in Ref.~\cite{Quaglioni2009}. In this section, we briefly outline the formalism and introduce notation relevant for the treatment of $3N$ interactions, which will be discussed in detail later.

The starting point are the {\it ab initio} NCSM eigenstates of the $(A{-}a)$-nucleon target and $a$-nucleon projectile  ($a<A$), which are used as building blocks of the $A$-body Hilbert space. Bound, resonant and scattering states of the $A$-nucleon system are all described within the RGM-inspired cluster basis
\begin{align}
|\Phi^{J^\pi T}_{\nu r}\rangle =& \Big [ \big ( \left|A{-}a\, \alpha_1 I_1^{\,\pi_1} T_1\right\rangle \left |a\,\alpha_2 I_2^{\,\pi_2} T_2\right\rangle\big ) ^{(s T)} \nonumber\\
& \times \,Y_{\ell}\left(\hat r_{A{-}a,a}\right)\Big ]^{(J^\pi T)}\,\frac{\delta(r-r_{A{-}a,a})}{rr_{A{-}a,a}}\,, \label{eq:basis}
\end{align}
where $ | A{-}a \, \alpha_1 I_1^{\,\pi_1} T_1 \rangle $ and $ |a \,\alpha_2 I_2^{\,\pi_2} T_2 \rangle$ denote translational-invariant  NCSM eigenstates in coordinate-space representation of the target and projectile, respectively, characterized by their energy labels $\alpha_i$, angular momentum-parity $I_i^{\pi_i}$, and isospin $T_i$ with $i{=}1,2$ the cluster index.
These translational-invariant cluster basis states describe two nuclei whose centers of mass are separated by the relative coordinate $\vec r_{A{-}a,a}$ and that are traveling in a $^{2s+1}\ell_J$ partial wave of relative motion (with $s$ the channel spin, $\ell$ the relative orbital angular momentum, and $J$ the total angular momentum of the system). To label the cluster basis states, we collect under the index $\nu$ the quantum numbers $\{A{-}a\,\alpha_1I_1^{\,\pi_1} T_1;\, a\, \alpha_2 I_2^{\,\pi_2} T_2;\, s\ell\}$ while additional quantum numbers are the parity $\pi=\pi_1\pi_2(-1)^{\ell}$ and the total isospin $T$. Finally, the antisymmetrization of the channel states of Eq.~(\ref{eq:basis}) with respect to exchange of nucleons pertaining to different clusters is enforced by the appropriate inter-cluster antisymmetrizer ${ { \mathcal  A}}_{\nu}$
\begin{align}
{ { \mathcal  A}}_{\nu} &= \sqrt{\frac{(A{-}a)!a!}{A!}} ~ \sum_{\sigma} \Pi_{\sigma} {\mathcal{P}}_\sigma \, ,
\end{align}
where the sum runs over all permutations $ {\mathcal{P}}_\sigma$ that exchange nucleons belonging to different clusters, while $\Pi_{\sigma}$ is the signature of the permutation.

The many-body wave function of the $A$-nucleon system is given in terms of the continuous basis set of Eq.~(\ref{eq:basis}) as:
\begin{align}
|\Psi^{J^\pi T}\rangle = &\sum_{\nu}\int dr ~ r^{2}  ~   {\mathcal A}_{\nu}|\Phi^{J^\pi T}_{\nu r}\rangle\,  \nonumber \\
\times &\sum_{\nu^\prime}\int dr^\prime r^{\prime\, 2} [{\cal N}^{-\frac12}]^{J^\pi T}_{\nu\nu^\prime}(r,r^\prime)\frac{\chi^{J^\pi T}_{\nu^\prime}(r^\prime)}{r^\prime}  \, , \label{eq:ansatz}
\end{align}
where $\chi^{J^\pi T}_{\nu'}(r')$ represent continuous linear variational amplitudes. These are determined by solving the orthogonalized NCSM/RGM equations (see, e.g., Sec.\ II.E of Ref.~\cite{Quaglioni2009}):
\begin{align}
\sum_{\gamma\gamma'\nu'} \! \int \! & dydy'dr^\prime y^2 y'^2 r^{\prime\,2} [{\mathcal N}^{-\frac12}]^{J^\pi T}_{\nu\gamma}(r,y){\mathcal H}^{J^\pi T}_{\gamma\gamma'}(y,y') \nonumber \\
\times & [\,{\mathcal N}^{-\frac12}]^{J^\pi T}_{\gamma'\nu'\,}(y',r') \frac{\chi^{J^\pi T}_{\nu^\prime} (r^\prime)}{r^\prime}= E\,\frac{\chi^{J^\pi T}_{\nu} (r)}{r}  \, , \label{eq:RGMeq}
\end{align}
where $E$ is the total energy in the center of mass (c.m.) frame. The norm and Hamiltonian integration kernels, $ {\mathcal N}^{J^\pi T}_{\nu\nu^\prime\,}(r,r^\prime)$ and ${\mathcal H}^{J^\pi T}_{\nu\nu^\prime\,}(r,r^\prime)$ respectively, are the  overlap and matrix elements of the Hamiltonian with respect to the antisymmetrized basis (\ref{eq:basis}), i.e.
\begin{align}
{\mathcal N}^{J^\pi T}_{\nu^\prime \nu}(r^\prime, r) &=  \langle \Phi^{J^\pi T}_{\nu^\prime r^\prime} | {\mathcal A}_{\nu^\prime} {\mathcal A}_{\nu} | \Phi^{J^\pi T}_{\nu r} \rangle\, , \nonumber \\
{\mathcal H}^{J^\pi T}_{\nu^\prime \nu}(r^\prime, r) &=  \langle\Phi^{J^\pi T}_{\nu^\prime r^\prime} |  {\mathcal A}_{\nu^\prime}  H  {\mathcal A}_{\nu} | \Phi^{J^\pi T}_{\nu r} \rangle \, ,
\label{eq:NH-kernel}
\end{align}
where $  H$ is
the microscopic $A$-nucleon Hamiltonian: 
\begin{equation}
  H =   T_{\rm rel}(r) +   {\mathcal V}_{\rm rel} +\bar{V}_{C}(r) +   H_{(A{-}a)} +   H_{(a)} \, .\label{eq:Hamiltonian}
\end{equation}
Here, $  H_{(A{-}a)}$ and $  H_{(a)}$ represent the $(A{-}a)$- and $a$-nucleon intrinsic Hamiltonians, $  T_{\rm rel}(r)$ is the relative kinetic energy and $\bar{V}_{\rm C}(r)=Z_{1\nu}Z_{2\nu}e^2/r$ is the average Coulomb interaction between pairs of clusters (with $Z_{1\nu}$ and $Z_{2\nu}$ the charge number of the clusters in channel $\nu$), and $  {\mathcal V}_{\rm rel}$ is the localized inter-cluster potential containing the microscopic nuclear interaction between nucleons in different clusters, given as:
\begin{align}
  {\mathcal V}_{\rm rel} =& \sum_{i=1}^{A{-}a}\sum_{j=A{-}a+1}^A {V}^{NN}_{ij} + \sum_{i<j=1}^{A{-}a}\sum_{k=A{-}a+1}^A {V}^{3N}_{ijk} \nonumber\\
&+ \sum_{i=1}^{A{-}a}\sum_{j<k=A{-}a+1}^A {V}^{3N}_{ijk} - \bar{V}_{\rm C}(r)\label{pot}\, .
\end{align}
In this work, $ {V}^{NN}$ and $ {V}^{3N}$ denote the nuclear plus Coulomb nucleon-nucleon and three-nucleon interactions, respectively. The same $NN$ and $3N$ interactions are used in both the $(A{-}a)$- and $a$-nucleon intrinsic Hamiltonians and in the inter-cluster potential $  {\mathcal V}_{\rm rel}$. 

The NCSM/RGM kernels are calculated from two main ingredients: $(i)$ the eigenstates of the projectile and target, and $(ii)$ the realistic nuclear interaction among the nucleons. The clusters' eigenstates are obtained by the diagonalization of $  H_{(A{-}a)}$ and $  H_{(a)}$ within the NCSM approach. This method employs a harmonic oscillator (HO) basis defined by the maximum number $N_{\rm max}$ of quanta above the unperturbed Slater determinant (SD) and the frequency denoted by  $\Omega$. The same HO frequency and consistent model-space sizes  are used for both the clusters' wave functions as well as for the localized parts of the integration kernels. The size $N_{\rm max}$ of the HO model space is the same for states of identical parity, whereas it differs by one unit for states of opposite parity.


\subsection{\label{kernels} Expression of the norm and Hamiltonian kernels in the Slater Determinant basis}

Before introducing the contributions to the Hamiltonian kernel due to the $3N$ component of the interaction, in this section we briefly review the main steps involved in the derivation of the integration kernels when the NCSM eigenstate of the target nucleus is expanded over a Slater-determinant basis.  For more details regarding these derivations we refer the interested reader to Secs.~II.C of Ref.~\cite{Quaglioni2009}.
 
Starting from Eqs.~(\ref{eq:NH-kernel}) and (\ref{eq:Hamiltonian}) it follows that the norm and Hamiltonian kernels may be written as
\begin{align}
{\mathcal N}^{J^\pi T}_{\nu^\prime\nu}(r^\prime, r)
&= \delta_{\nu^\prime\nu}\frac{\delta(r^\prime-r)}{r^\prime r} + {\mathcal N}^{\rm ex}_{\nu^\prime\nu}(r^\prime, r) \label{eq:N-kernel-2}
\end{align}
and,
\begin{align}
{\mathcal H}^{J^\pi T}_{\nu^\prime\nu}(r^\prime, r) = & \langle \Phi^{J^\pi T}_{\nu^\prime r^\prime} | \frac12( {\mathcal A}_{\nu^\prime} {\mathcal A}_{\nu}  H+  H {\mathcal A}_{\nu^\prime} {\mathcal A}_{\nu}) | \Phi^{J^\pi T}_{\nu r} \rangle \nonumber\\
= & \frac12 \left \{ \big[ {T}_{\rm rel}(r')+\bar{V}_C(r')+E_{\alpha_1'}^{I_1'T_1'} +E_{\alpha_2'}^{I_2'T_2'} \big] \right .\nonumber \\
\, &\times \left . \mathcal{N}_{\nu'\nu}^{J^\pi T}(r', r)+\mathcal{V}^{J^\pi T}_{\nu' \nu}(r',r) \right \} + \frac12\{h.c.\}\, , \label{eq:H-kernel-2}
\end{align}
where $\{h.c.\}$ stands for the hermitian conjugate term coming from the $  H {\mathcal A}_{\nu^\prime} {\mathcal A}_{\nu}$ operator, the energy eigenvalues of the target and projectile are given as $E_{\alpha_i'}^{I_i'T_i'}$, and the exchange part of the norm kernel and the potential kernel are denoted respectively by ${\mathcal N}^{\rm ex}_{\nu^\prime\nu}(r^\prime, r)$  and $\mathcal{V}^{J^\pi T}_{\nu' \nu}(r',r)$. These integration kernels are obtained in the NCSM/RGM model space by expanding the  Dirac delta function of Eq.~(\ref{eq:basis}) on a set of HO radial wave functions with frequency $\Omega$ identical to that used for the two clusters. This leads to:
\begin{align}
{\mathcal N}^{\rm ex}_{\nu^\prime\nu}(r^\prime, r) =& \sum_{n^\prime n}R_{n^\prime\ell^\prime}(r^\prime)R_{n\ell}(r)  \nonumber\\
&\times \langle\Phi^{J^\pi T}_{\nu^\prime n^\prime} | \sqrt{\tfrac{ A! }{ (A{-}a^\prime)! a^\prime !}}   {\mathcal A}_\nu  |\Phi^{J^\pi T}_{\nu n} \rangle \nonumber\\
=&\sum_{n^\prime n}R_{n^\prime\ell^\prime}(r^\prime)R_{n\ell}(r) ~{\mathcal N}^{\rm ex}_{\nu^\prime n'\nu n}
\label{eq:norm-modspace}
\end{align}
and 
\begin{align}
\mathcal{V}^{J^\pi T}_{\nu' \nu}(r',r) =& \sum_{n^\prime n}R_{n^\prime\ell^\prime}(r^\prime)R_{n\ell}(r) \nonumber \\
&\times \langle\Phi^{J^\pi T}_{\nu^\prime n^\prime} | \sqrt{\tfrac{ A!}{ (A{-}a^\prime)! a^\prime !}} { {\mathcal V}}_{\rm rel} {\mathcal A}_{\nu} |\Phi^{J^\pi T}_{\nu n} \rangle \nonumber\\
=&\sum_{n^\prime n}R_{n^\prime\ell^\prime}(r^\prime)R_{n\ell}(r) ~\mathcal{V}^{J^\pi T}_{\nu^\prime n'\nu n} \,, \label{eq:pot-kernel}
\end{align}
where we have introduced the HO channel states
\begin{align}
|\Phi^{J^\pi T}_{\nu n}\rangle =& \Big [ \big ( \left|A{-}a\, \alpha_1 I_1^{\,\pi_1} T_1\right\rangle \left |a\,\alpha_2 I_2^{\,\pi_2} T_2\right\rangle\big ) ^{(s T)} \nonumber \\
&\times Y_{\ell}\left(\hat r_{A{-}a,a}\right)\Big ]^{(J^\pi T)}\,R_{n\ell}(r_{A{-}a,a})\,,\label{eq:ho-basis-n}
\end{align}
and the model-space kernels ${\mathcal N}^{\rm ex}_{\nu^\prime n'\nu n}$ and $\mathcal{V}^{J^\pi T}_{\nu^\prime n'\nu n}$. In deriving these expressions, we have taken advantage of the commutation relationship between the inter-cluster antisymmetrizers and the microscopic Hamiltonian, $[ {\mathcal A}_{\nu},  H]=0$, and the fact that $ {\mathcal A}_{\nu^\prime} {\mathcal A}_{\nu} |\Phi^{J^\pi T}_{\nu n}\rangle = \sqrt{\tfrac{ A!}{ (A{-}a^\prime)! a^\prime !}}  {\mathcal A}_{\nu}|\Phi^{J^\pi T}_{\nu n}\rangle$.
The HO radial quantum numbers in Eqs.~(\ref{eq:norm-modspace}) and (\ref{eq:pot-kernel}) run up to the maximum value of $N^{\rm RGM}_{\rm max}$ which is chosen consistently with the model space of the targets and projectiles. Such an expansion in HO radial wave functions is accurate and justified for matrix elements of localized operators such as those entering the exchange part of the norm, and the potential kernel.

While the translationally-invariant channel states of Eq.~(\ref{eq:ho-basis-n}) represent a ``natural" choice for the calculation of the integration kernels, from a computational point of view it is convenient to work with NCSM eigenstates of the $(A{-}a)$-nucleon cluster expanded in a many-body space of HO Slater determinants, 
\begin{align}
 |A{-}a\, \alpha_1 I_1^{\pi_1}&M_1 T_1 M_{T_1} \rangle_{\rm SD} \label{eq:SD-eigenstate} \\
 &=\;  |A{-}a\, \alpha_1 I_1^{\pi_1}M_1 T_1 M_{T_1} \;\rangle\,\varphi_{000}(\vec R^{(A{-}a)}_{\rm c.m.}) \,,\nonumber 
\end{align}  
and to introduce the SD channel states
\begin{eqnarray}
|\Phi^{J^\pi T}_{\nu n}\rangle_{\rm SD}   &=&    \Big [\big ( |A{-}a\, \alpha_1 I_1^{\pi_1} T_1 \rangle_{\rm SD} 
\left |a\,\alpha_2 I_2^{\pi_2} T_2\right\rangle\big )^{(s T)}\nonumber\\
&&\times Y_{\ell}(\hat R^{(a)}_{\rm c.m.})\Big ]^{(J^\pi T)} R_{n\ell}(R^{(a)}_{\rm c.m.})\, .  \label{eq:SD-basis}
\end{eqnarray}
Here $R^{(a)}_{\rm c.m.}$ denotes the center of mass of the $a$-nucleon cluster and 
$\varphi_{000}(\vec R^{(A{-}a)}_{\rm c.m.})=R_{00}(R^{(A-a)}_{\rm c.m.})Y_{00}(\hat R^{(A{-}a)}_{\rm c.m.})$ represents the motion of the target c.m., which factorizes exactly in a NCSM calculation due to the $N_{\rm max}$ truncation scheme \cite{Barrett2013}. Correspondingly, the matrix elements of any translational-invariant operator with respect to the SD channel states of Eq.~(\ref{eq:SD-basis}) are affected by components of the c.m.\ of the overall system. However, this c.m.~motion can be removed [and the translationally-invariant matrix elements with respect to the channel states (\ref{eq:ho-basis-n}) recovered] exactly by inverting a linear transformation as described in Sec.\ II.C.2 of Ref.~\cite{Quaglioni2009}. 

In the following, we present the contributions of the $3N$ interaction to the potential kernel $\mathcal{V}^{J^\pi T}_{\nu^\prime n'\nu n}$ as defined in Eq.~\eqref{eq:pot-kernel} for the description of nucleon-nucleus collisions. That is, we consider the case of projectile-target binary-cluster channels of both initial and final states that are formed by a single nucleon ($a=a^\prime=1$) in relative motion with respect to a target of $A{-}1$ nucleons. Then, the inter-cluster antisymmetrizer simplifies to
\begin{eqnarray}
   {\mathcal{A}}_{\nu} & \equiv  {\mathcal{A}} =&
      \frac{1}{\sqrt{A}}\Big( 1-\sum_{i=1}^{A-1} {P}_{iA}\Big),
      \label{eq:antisym1}
\end{eqnarray} 
where $ {P}_{i,A}$ is the transposition operator exchanging the $i$-th particle in the target with the projectile nucleon, labelled by the index $A$. 
In addition, the calculation of matrix elements in the basis~(\ref{eq:SD-basis}) is most efficiently achieved by first performing the transformation
\begin{align}
|\Phi^{J^\pi T}_{\nu n}\rangle_{\rm SD}   =& \sum_j ~ \hat{s}\hat{j} ~ (-1)^{I_1+J+j} \left\{ \begin{array}{@{\!~}c@{\!~}c@{\!~}c@{\!~}} I_1 & \frac{1}{2} & s \\[2mm] \ell & J & j \end{array}\right\} |\Phi^{J^\pi T}_{\kappa}\rangle_{\rm SD}\, , \label{SD-basis-SNP}
\end{align}
to the new SD basis labeled by the cumulative index $\kappa=\{A{-}1\,\alpha_1 I_1^{\pi_1} T_1; n \ell j \tfrac12\}$
\begin{align}
|\Phi^{J^\pi T}_{\kappa}\rangle_{\rm SD} &= \Big[\left|A{-}1\, \alpha_1 I_1^{\pi_1} T_1\right\rangle_{\rm SD} |n \ell j \tfrac12\rangle \Big]^{(J^\pi T)} \notag \\
&= \sum_{M_1 m_j} \sum_{M_{T_1} m_t} \clebsch{I_1}{j}{J}{M_{1}}{m_j}{M_J} \clebsch{T_1}{\tfrac{1}{2}}{T}{M_{T_1}}{m_t}{M_T}\notag \\
&\times |A{-}1\, \alpha_1 I_1^{\pi_1} M_1 T_1 M_{T_1}\rangle_{\rm{SD}} |n\ell jm_j\tfrac{1}{2}m_t\rangle \label{eq:new-basis-formula}
\end{align}
where $\hat{x}=\sqrt{2x+1}$, and $|n\ell jm_j\tfrac{1}{2}m_t\rangle$ denotes the HO single-particle state of the projectile. In this new basis the target is expressed in terms of Slater-determinants, thereby we can use powerful second quantization techniques to calculate the potential kernels. 


\subsection{\label{NNN-kernels} Three-nucleon potential kernel}
In this section, we derive 
the contributions of the $3N$ interaction to the potential kernel. After presenting some of the resulting algebraic expressions, we discuss in detail two alternative implementations of the formalism, which have been used to benchmark the present nucleon-$^4$He calculations.

\begin{figure}[t]
\includegraphics*[width=0.85\columnwidth]{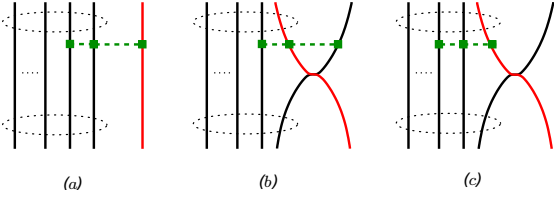}
\caption{Diagrammatic representation of the direct ($(a)$ and $(b)$) and exchange components of the $3N$ potential kernel. The groups of circled lines represent the $(A{-}1)$-nucleon cluster. Bottom and upper part of the diagram represent initial and final states, respectively.}\label{diagram-NNNpot}
\end{figure}

Starting from Eq.~(\ref{eq:pot-kernel}) the inclusion of the $3N$ force into the single-nucleon projectile formalism, although quite involved, is straightforward. Using the expression~\eqref{eq:antisym1} we obtain two new contributions
\begin{widetext}
\begin{align}
{{\mathcal V}^{3N}_{\nu^\prime n'\nu n}}= & \phantom{-} \frac{(A{-}1)(A{-}2)}{2} ~~ \leftsub{\rm SD}{\langle}\Phi^{J^\pi T}_{\nu^\prime n^\prime} |  {V}^{3N}_{A-2,A-1,A}(1-2  P_{A-1,A}) | \Phi^{J^\pi T}_{\nu n}\rangle_{\rm SD} \label{pot-NNN-direct} \\
&- \frac{(A{-}1)(A{-}2)(A{-}3)}{2} ~~ \leftsub{\rm SD}{\langle}\Phi^{J^\pi T}_{\nu^\prime n^\prime} |   P_{A-1,A} {V}^{3N}_{A-3,A-2,A-1} |\Phi^{J^\pi T}_{\nu n} \rangle_{\rm SD}\, . \label{pot-NNN-ex}
\end{align}
\end{widetext}
As for the corresponding $NN$ portion of the nucleon-nucleus potential kernel presented in Ref.~\cite{Quaglioni2009}, we identify a direct [Eq.~\eqref{pot-NNN-direct}] and an exchange [Eq.~\eqref{pot-NNN-ex}] term, described by diagrams $(a)$ and $(b)$, and diagram $(c)$ of Fig.~\ref{diagram-NNNpot}, respectively.

As illustrated in Fig.~\ref{diagram-NNNpot}, the direct and exchange terms involve operations on two and three nucleons of the target, respectively. Hence, they require the evaluation of matrix elements of two- and three-body densities of the target, respectively. In comparison, the $NN$ potential kernel for $(A{-}1,1)$ binary-cluster channels in both initial and final states depends only on matrix elements of one- and two-body densities of the target. This highlights the increased computational cost of including the $3N$ interaction in the NCSM/RGM approach.

In the following, we focus in detail on the derivation of the exchange potential kernel of Eq.~\eqref{pot-NNN-ex}, while leaving the algebraic expression of the $3N$ contribution to the direct kernel to the Appendix~\ref{sec:appendix-direct}. 
To derive the expression for the exchange kernel we compute the matrix elements of $  P_{A-1,A} {V}_{A-3,A-2,A-1}$ with respect to the basis states~\eqref{eq:new-basis-formula}.\ We insert identity operators in terms of $A$-body HO Slater-determinants, explicitly perform the transposition, introduce antisymmetrized HO matrix elements of the $3N$ force by exploiting the fact that $|\Phi^{J^\pi T}_{\kappa}\rangle_{\rm SD}$ is antisymmetric with respect to the exchange of nucleons belonging to the target and obtain
\begin{widetext}
\begin{align}
\leftsub{\rm SD}{\langle}&\Phi^{J^\pi T}_{\kappa^\prime}\big| P_{A-1,A}  V^{3N}_{A-3\,A-2\,A-1} \ket{\Phi^{J^\pi T}_{\kappa}}_{\rm SD} = \frac{1}{6(A-1)(A-2)(A-3)}\sum_{M_1 m_j} \sum_{M_{T_1} m_t}\sum_{M^{\prime}_1 m^{\prime}_j} \sum_{M^{\prime}_{T_1} m^{\prime}_t}  \notag \\
\times&   \clebsch{I_1}{j}{J}{M_{1}}{m_j}{M_J} \clebsch{T_1}{\tfrac{1}{2}}{T}{M_{T_1}}{m_t}{M_T} 
\clebsch{I^{\prime}_1}{j^{\prime}}{J}{M^{\prime}_{1}}{m^{\prime}_j}{M^{\prime}_J} \clebsch{T^{\prime}_1}{\tfrac{1}{2}}{T}{M^{\prime}_{T_1}}{m^{\prime}_t}{M^{\prime}_T}  \notag \\
\times& \sum_{abdef}\bra{ab\;n^{\prime}\ell^{\prime}j^{\prime}m_{j}^{\prime}\tfrac{1}{2}m^{\prime}_{t}}V^{3N}\ket{def} \;
  \leftsub{\rm SD}{\bra{A{-}1\, \alpha_1^{\prime} I_1'^{\pi_1^{\prime}} M_1^{\prime} T_1^{\prime} M_{T_1}^{\prime}}} a^{\dagger}_{n\ell jm_{j}\tfrac{1}{2}m_{t}} a^{\dagger}_{b} a^{\dagger}_{a} 
a_{d} a_{e} a_{f} \ket{ A{-}1\, \alpha_1 I_1^{\pi_1} M_1 T_1 M_{T_1}}_{\rm SD} \, , \label{NNN-ex-express}
\end{align}
\end{widetext}
where we introduced  the notation $\{a,\ldots, f\}$ denoting the quantum numbers of $\ell s$-coupled HO single-particle states, i.e. $a  = \{n_a \ell_a j_a m_{j_a} m_{t_a}\}$. From a numerical perspective, we have at this point two options: ({\em{i}}) implementing Eq.~\eqref{NNN-ex-express} directly, or ({\em ii}) introducing coupled densities and further algebraic manipulations as previously done for the $NN$ case \cite{Quaglioni2009}. We pursued both options and will discuss these two alternative approaches in detail later on.
To arrive at the coupled densities, we make use of the $JT$-coupled $3N$ matrix elements as introduced in Refs.~\cite{Roth2011} and~\cite{Roth2013}, and use reduced matrix elements of the target density by means of the Wigner-Eckart theorem. Accordingly, we evaluate all summations over projection quantum numbers and arrive at
\begin{widetext}
\begin{align}
\leftsub{\rm SD}{\langle} & \Phi^{J^\pi T}_{\kappa^\prime} |   P_{A-1,A}V^{3N}_{A-3,A-2,A-1}  |\Phi^{J^\pi T}_{\kappa}\rangle_{\rm SD}  =  \frac{1}{6(A{-}1)(A{-}2)(A{-}3)} \sum_{{\bar a} {\bar b} {\bar d} {\bar e} {\bar f}} \sum_{J_{ab} T_{ab}} \sum_{ J_0 T_0 } \sum_{J_{de} T_{de}} \sum_{ J_g T_g} \sum_{K \tau}\nonumber\\
\times & {\hat J_0}{\hat T_0} {\hat J_g}{\hat T_g}{\hat K}{\hat \tau}  ~ (-1)^{j+j'+J_{ab}+K-J_g+I_1+J + 1+T_{ab}+\tau-T_g+T_1+T} \langle {\bar a}, {\bar b}  ; J_{ab} T_{ab},{n'\ell 'j'\tfrac{1}{2}} ; J_0 T_0 | V^{3N}  |  {\bar d}, {\bar e}  ; J_{de} T_{de},{\bar f} ; J_0 T_0 \rangle \nonumber\\
\times& \left\{ 
\begin{array}{ccc} 
I_1 &K&I_1' \\
 j'&J&j 
\end{array} \right\}
\left\{
\begin{array}{ccc}
j'&K&j \\
J_g&J_{ab}&J_0
\end{array}
\right\}
\left\{
\begin{array}{ccc}
T_1 &\tau&T_1'\\
\tfrac12 &T& \tfrac12
\end{array}
\right\}
\left\{
\begin{array}{ccc}
\tfrac12 &\tau& \tfrac12\\
T_g&T_{ab}&T_0
\end{array}
\right\}  \nonumber\\
\times &\leftsub{\rm SD}{\langle} A{-}1 \alpha_{1}' I_1^{\prime\,\pi_1'} T_1' ~|||~  \Big( \Big( \big( a^{\dag}_{{\bar a}}  a^{\dag}_{{\bar b}} \big)^{J_{ab} T_{ab}} a^{\dag}_{n\ell j \tfrac{1}{2} } \Big)^{J_g T_g} \Big( \big( {\tilde a}_{{\bar d}}{\tilde a}_{{\bar e}} \big)^{J_{de}T_{de}}  {\tilde a}_{{\bar f}} \Big)^{J_0 T_0} \Big)^{K \tau} ~|||~A{-}1 \alpha_{1} I_1^{\pi_1} T_1 \rangle_{\rm SD}  \; ,
\label{NNN-ex-express2}
\end{align}
\end{widetext} 
where $\tilde{a}_{f}$ is the time-reversed annihilation operator 
\begin{align}
\tilde{a}_{f}=(-1)^{j_{f}-m_{j_f}+\tfrac12-m_{t_{f}}}\;a_{n_{f} \ell_{f} j_{f} -m_{j_f} -m_{t_{f}}}\, , \nonumber
\end{align}
$\{{\bar a},\ldots,{\bar f}\}$ denote HO orbitals, i.e. ${\bar a }= \{n_{ f}, \ell_{ f}, j_{ f}\}$, and the triple vertical bars indicate that the matrix elements are reduced in both angular momentum and isospin.

To avoid the explicit treatment of the three-body density matrix, we factorize this expression by inserting a completeness relationship over $(A{-}4)$-body eigenstates leading to 
\begin{widetext}
\begin{align}
\leftsub{\rm SD}{\langle} & \Phi^{J^\pi T}_{\kappa^\prime} |   P_{A-1,A}V^{3N}_{A-3,A-2,A-1}  |\Phi^{J^\pi T}_{\kappa}\rangle_{\rm SD}  =  \frac{1}{6(A{-}1)(A{-}2)(A{-}3)} \sum_{{\bar a} {\bar b} {\bar d} {\bar e} {\bar f}} \sum_{J_{ab} T_{ab}} \sum_{ J_0 T_0 } \sum_{J_{de} T_{de}} \sum_{ J_g T_g} \sum_{ \beta} \nonumber\\
\times & {\hat J_0}{\hat T_0} {\hat J_g}{\hat T_g}   ~ \left\{ \begin{array}{ccc} I_{\beta} &J_g &I_1' \\ J_0&J_{ab}& j' \\ I_1&j&J \end{array} \right\}   \left\{ \begin{array}{ccc} T_{\beta} &T_g&T_1' \\ T_0&T_{ab}& \tfrac12  \\ T_1&\tfrac12&T \end{array} \right\} 
\leftsub{\rm SD}{\langle} A{-}1\alpha_{1}' I_1^{\prime\,\pi_1'} T_1' ||| \Big( \big(a^{\dag}_{{\bar a}}  a^{\dag}_{{\bar b}} \big)^{J_{ab} T_{ab}}a^{\dag}_{n\ell j \tfrac{1}{2} }\Big)^{J_g T_g} ||| A{-}4 \alpha_{\beta} I_{\beta}^{\pi_\beta} T_{\beta} \rangle_{\rm SD} \nonumber \\
\times&\langle {\bar a}, {\bar b}  ; J_{ab} T_{ab},n'\ell' j' \tfrac{1}{2} ; J_0 T_0 |  V^{3N}  |  {\bar d}, {\bar e}  ; J_{de} T_{de},{\bar f} ; J_0 T_0 \rangle \leftsub{\rm SD}{\langle} A{-}1\alpha_{1} I_1^{\pi_1} T_1 ||| \Big( \big( {a}^{\dag}_{{\bar d}} {a}^{\dag}_{{\bar e}} \big)^{J_{de}T_{de}}  {a}^{\dag}_{{\bar f}} \Big)^{J_0T_0}  ||| A{-}4\alpha_{\beta}I_{\beta}^{\pi_\beta} T_{\beta} \rangle_{\rm SD} \; . \label{NNN-ex-express3}
\end{align}
\end{widetext}
Compared to Eq.~(\ref{NNN-ex-express2}), there is an additional summation over the index $\beta$ labeling the eigenstates $| A{-}4 \alpha_{\beta},I_{\beta}^{\pi_\beta},T_{\beta} \rangle$ of the $(A{-}4)$-body system.  For the case of nucleon-$^4$He scattering, which will be the focus of Sec.~\ref{sec:applications}, they consist of single-particle states.

Equations~(\ref{NNN-ex-express}) and~(\ref{NNN-ex-express3}) summarize the two approaches ({\em i}) and ({\em ii}) we have developed to calculate the potential kernels. We used these two implementations to benchmark our results.
In the first approach ({\em i}) according to Eq.~\eqref{NNN-ex-express}, we have to deal with the three-body density explicitly. Storing in memory three-body density matrices is of course very demanding. Therefore, we have developed an efficient on-the-fly computation of these matrix elements by exploiting the fact that the target eigenstates $|A{-}1\, \alpha_1 I_1^{\pi_1} M_1 T_1 M_{T_1}\rangle$ are implicitly given as expansion in HO many-body SDs within the NCSM model space. We can pull the two summations corresponding to these expansions in front of all other summations in Eq.~\eqref{NNN-ex-express}. In this way we obtain an expression in which each term can be computed independently, i.e., it is ideally suited for parallel computation with perfect scaling. In addition, the sums over HO single-particle states are of course restricted to those combinations, which can connect the two SDs of the density matrix. Here, we can make use of the technology that was originally developed to compute $A$-body matrix elements of three-body operators during the setup of the many-body matrix in the importance-truncated NCSM \cite{Roth2007,Roth2009}. The next critical objects in Eq.~\eqref{NNN-ex-express} are the $m$-scheme matrix elements of the $3N$ interaction. The storage of these matrix elements in memory is again prohibitive if we want to proceed to large model spaces. However, we benefit from storing the matrix elements of the $3N$ interaction in the $JT$-coupled scheme developed by Roth et al.~\cite{Roth2011, Roth2013} and the corresponding efficient on-the-fly decoupling into the $m$-scheme. Finally, we note from Eq.~\eqref{NNN-ex-express} the necessity to treat the projection quantum numbers of the angular momenta and isospins of the target states explicitly, including consistent relative phases. Both can be accomplished using a single NCSM run to produce a specific eigenvector from which all other vectors with necessary projection are obtained using angular momentum raising and lowering operators.

In the second approach ({\em ii}), we first calculate and store in memory the reduced matrix elements of the tensor operator
\begin{align}
\Big( \big( {a}^{\dag}_{{\bar d}} {a}^{\dag}_{{\bar e}} \big)^{J_{de}T_{de}}  {a}^{\dag}_{{\bar f}} \Big)^{J_0T_0} \nonumber
\end{align}
and compute the factorized three-body density of Eq.~(\ref{NNN-ex-express3}) on the fly. This strategy reduces the computational burden and computer memory required to perform the calculation. We work directly with the $JT$-coupled $3N$ matrix elements exploiting their symmetries and using the appropriate Racah algebra if necessary. The main limitation of this approach is the factorization of the reduced density which is feasible only for light systems where a complete set of $(A{-}4)$-body eigenvectors can be obtained, i.e., the four- and five-nucleon systems for the specific case of nucleon-nucleus collisions. For such systems, however, it is still a more efficient approach when many excited states of the target are included in the calculation as in the present results with seven \elem{He}{4} eigenstates. In terms of numerics we have achieved a load-balanced parallel implementation by using a non-blocking master-slave algorithm. Finally, following the application of the Wigner-Eckart theorem, the isospin dependence of the nuclear Hamiltonian in the potential kernels cannot  be treated exactly in this approach. To account for the breaking of isospin symmetry of the $NN$ interaction we determine the coefficients of the combination of $nn$, $pp$ and $np$ parts of the $NN$ potential with the help of the Hamiltonian kernel expressed in spirit of approach ({\em i}) that treats exactly the isospin-symmetry breaking. This analysis will be discussed in further detail in Sec.~\ref{sec:isospin_average}, where it is illustrated and benchmarked in the $p$-$^4$He scattering problem. 

With the exception of the treatment of isospin symmetry, the two implementations described in this section are formally equivalent. By storing in memory the reduced densities, the latter is more efficient for reactions with different projectiles while the former is ideally suited for addressing heavier targets.


\section{Applications to nucleon-$^4$He scattering}\label{sec:applications}
Nucleon-$^4$He scattering has been an important testing ground for {\em ab initio} reaction calculations. 
The tightly-bound nature of the $^4$He nucleus makes it one of the simplest scattering systems to be described, characterized by a single open (binary-cluster) channel up to fairly high energies, and hence an important benchmark for any approach to light-nucleus reactions.  At the same time, its sensitivity to the strength of the spin-orbit force, as seen in earlier {\em ab initio} studies of the $^{2}P_{3/2}$ and $^{2}P_{1/2}$ scattering phase shifts~\cite{Nollett2007,Quaglioni2008,Quaglioni2009}, makes it an ideal stage for the investigation of $3N$-force effects in the continuum.  

The role of the $3N$ force in the description of the spin-orbit splitting between $^{2}P_{3/2}$ and $^{2}P_{1/2}$ resonances was first investigated in Ref.~\cite{Nollett2007}, where these and the non-resonant $1/2^+$ phase shifts below 5\,MeV c.m.\ energy were obtained in GFMC calculations with the AV18 $NN$ \cite{Wiringa1995} plus UIX and IL2 $3N$ interaction models \cite{Pieper2001}. This work was followed soon after by NCSM/RGM calculations of $n$-$^4$He scattering for energies up to the inelastic threshold with several realistic $NN$ potentials~\cite{Quaglioni2008,Quaglioni2009}. Here, the insufficient splitting between the $^{2}P_{3/2}$ and $^{2}P_{1/2}$ resonances found using the chiral N$^3$LO $NN$ potential of Ref.~\cite{Entem2003} was attributed in part to the omission of the $3N$ terms of the chiral interaction. In the following we revisit these calculations by including, for the first time, the N$^2$LO $3N$ force of Ref.~\cite{Navratil2007}. 

Different from Refs.~\cite{Quaglioni2008,Quaglioni2009}, where the strong short-range correlations generated by the N$^3$LO $NN$ potential where treated by means of a modified Lee-Suzuki two-body effective interaction, here both chiral $NN$ and $3N$ forces are softened through an SRG transformation at the three-body level. We also note that the present study goes beyond that published in Ref.~\cite{Navratil2010}, where nucleon-$^4$He scattering results were obtained using only the two-body portion ($NN$-only) of the SRG-evolved chiral N$^3$LO $NN$ potential.

The discussion of the results is organized as follows: we start with some technical details about the chiral nuclear interactions used throughout this section, including a short description of the similarity renormalization group technique followed by a study of the convergence of our phase shifts with respect to the parameters of the NCSM/RGM model-space and of the effects of the isospin-symmetry breaking of the interaction. Finally, we conclude with an analysis of the effect of the initial $3N$ interaction on phase shifts, angular cross sections and polarization observables including the first seven low-lying states of the \elem{He}{4} target and compare our results to experiment.


\subsection{SRG-transformed $NN$+$3N$ Hamiltonian}\label{sec:Hamiltonian&SRG}

One cornerstone of an {\em ab initio} calculation is a high-precision nuclear Hamiltonian, which is in the ideal case solidly rooted in quantum chromodynamics. Throughout this study of nucleon-\elem{He}{4} scattering we employ the $NN$ interaction from $\chi$EFT at N$^3$LO by Entem and Machleidt \cite{Entem2003} along with the local form of the chiral $3N$ interaction at N$^2$LO \cite{Navratil2007}, both with cutoff 500 $\text{MeV}/c$. Note that the low-energy constants of this Hamiltonian are entirely determined from two- and three-body systems \cite{Gazit2009}. Currently, this is one of the best nuclear Hamiltonians available in terms of HO matrix elements, as needed here. We soften these interactions by a continuous unitary transformation of the $NN$ and $3N$ interaction using the similarity renormalization group technique \cite{Bogner2007,Jurgenson2009,Jurgenson2011,Roth2011, Roth2010}. With decreasing resolution scale $\lambda$ (or likewise with increasing flow parameter $\alpha=\lambda^{-4}$ as used elsewhere \cite{Roth2011, Roth2012}), this transformation tames the short-range correlations in the many-body eigenstates resulting in an improved convergence behavior with respect to the model-space size, both in the NCSM when computing the target eigenvectors as well as in the NCSM/RGM when computing the Hamiltonian integration kernel. However, the transformation formally induces irreducible many-body forces up to the $A$-body level. Today, the SRG transformation can be performed consistently in either the two- or the three-body space, allowing us to distinguish the effects of induced $3N$ forces from those of the initial $3N$ interactions \cite{Jurgenson2011,Roth2011}. Therefore, we define the three different Hamiltonians which we will investigate in the following: $NN$-only, resulting from the SRG transformation of the initial $NN$ interaction in two-body space (i.e., no $3N$-force effects are included); $NN$+$3N$-induced, resulting from the SRG transformation of the initial $NN$ interaction in three-body space (i.e., SRG-induced three-body contributions are taken into account);  $NN$+$3N$-full, resulting from the SRG evolution of the initial $NN$+$3N$ Hamiltonian in three-body space (i.e. already the initial Hamiltonian contains $3N$ components and all $3N$ terms are retained). By variation of the resolution scale $\lambda$ we can investigate the relevance of neglected  higher-order induced many-body forces in the context of scattering observables, as before in the case of nuclear structure observables (see, e.g., Refs.~\cite{Jurgenson2009} and~\cite{Roth2011}). For this variation we will use the SRG resolution scales (flow parameters) $\lambda=1.88\,\text{fm}^{-1}$ ($\alpha=0.08\,\text{fm}^4$), $\lambda=2.0\,\text{fm}^{-1}$ ($\alpha=0.0625\,\text{fm}^4$) and $\lambda=2.24\,\text{fm}^{-1}$ ($\alpha=0.04\,\text{fm}^4$). These values fall in a range for which the interaction is sufficiently soft to reach convergence within our largest model space of $N_{\text{max}} =13$ and, at the same time, four- and higher-order many-body forces have been found to be negligible in nuclear structure calculations for \elem{He}{4} \cite{Jurgenson2009, Roth2011}.

Concerning the $3N$ interaction in the NCSM/RGM calculations, we have to limit the sum of the HO single-particle energy quantum numbers by $e_a+e_b+e_c\leq E_{3\text{max}}$ as it was also discussed for other many-body methods~\cite{Roth2012, Binder2013, Hergert2013, Hergert2013a}. For the calculation of the NCSM eigenstates of the \elem{He}{4} target considered later on, this is not an issue because $E_{3\text{max}} {\geq} N_{\text{max}}$. However, the NCSM/RGM potential kernels for nucleon-nucleus collisions may be affected by this cut since in principle they probe $3N$ matrix elements up to $E_{3\text{max}}{=}2N_{\text{max}}$, as can be seen from Eq.~\eqref{NNN-ex-express} and the fact that the single-particle model space of the projectile has also to be truncated by $N_{\text{max}}$ for an exact recovery of the translational-invariant matrix elements from the SD ones (see Sec.~\ref{kernels}). We will check the sensitivity of our results with respect to $E_{3\text{max}}$ in Sec.~\ref{convergence}.  Unless otherwise stated, the matrix elements of the $3N$ force are included up to $E_{3\rm{max}}{=}14$.


\subsection{Convergence properties and approximations}
\label{convergence}
 
In this section we present a comprehensive study of all relevant parameter variations within the NCSM/RGM approach including explicit $3N$ interactions. We perform this analysis on the scattering phase shifts, which are obtained by solving the NCSM/RGM equations~(\ref{eq:RGMeq}) with scattering boundary conditions by means of the calculable $R$-matrix method on Lagrange mesh (see, e.g., Ref.~\cite{Descouvemont2010} and references therein). Note that in this work, the asymptotic limit of the two-body nuclear scattering problem is matched at the channel radius of $18\,\rm fm$, large enough for the clusters to interact only through the Coulomb force. We explicitly checked that all present results are independent of such parameter.

\subsubsection{Dependence on $N_{\rm max}$}

\begin{figure}[t]
\includegraphics*[width=1\columnwidth]{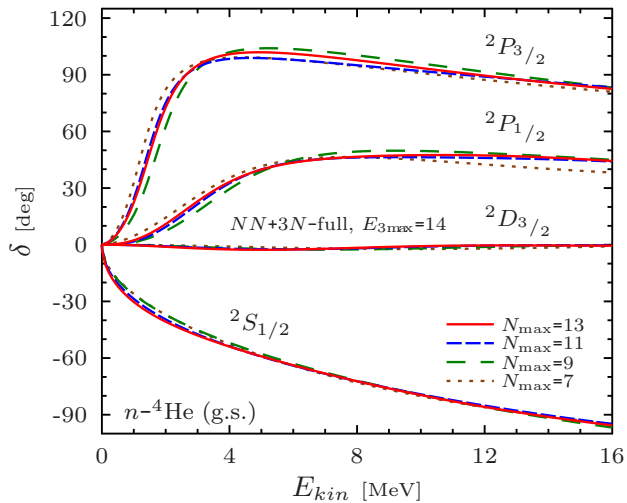}
\caption{
   (color online) Convergence of the $n$-$^4$He $^1S_{1/2}$, $^2P_{1/2}$, $^2P_{3/2}$, and $^2D_{3/2}$ phase shifts with respect to the model-space size $N_{\rm max}$. Brown dotted lines, green long-dashed lines, blue dashed lines and red solid lines correspond to $N_{\text{max}} = 7,9,11$ and 13, respectively. All curves were obtained including the g.s.\ of $^4$He and employing the SRG-evolved chiral $NN$+$3N$-full interaction with $\lambda=2.0$ fm$^{-1}$, $E_{3\rm{max}}{=}14$ and $\hbar\Omega{=}20$ MeV.} \label{fig:Nmax-convergence}
\end{figure}

We begin our study by exploring the convergence properties of the $n$-$^4$He phase shifts with respect to the model-space size $N_{\rm max}$ of the HO basis used to expand the localized parts of the NCSM/RGM kernels and NCSM wave functions of the target. These results were obtained with the SRG-evolved chiral $NN$+$3N$-full Hamiltonian with resolution scale $\lambda{=}2.0$ fm$^{-1}$ and an HO frequency of $\hbar\Omega{=}20$ MeV. Figure~\ref{fig:Nmax-convergence} presents single channel calculations for $N_{\rm max}{=}7,9,11$, and 13 carried out using $n$-$^4$He channel states with the $^4$He in its ground state (g.s.). The phase shifts for the first four partial waves exhibit a good convergence behavior. With the exception of the $^2P_{3/2}$ resonance, where we can observe a difference of less than 5 deg in the energy region $4\le E_{kin}\le 10$ MeV, the $ N_{\rm max}{=}11$ and $ N_{\rm max}{=}13$  phase shifts are on top of each other. An analogous behavior is obtained when using the $NN$+$3N$-induced interaction. Therefore, given the large scale of these computations, we performed some of the remaining sensitivity studies within an $N_{\rm max}=11$ model space. In addition, the convergence behavior presented in Fig.~\ref{fig:Nmax-convergence} is very similar to that obtained in Ref.~\cite{Navratil2010} for the N$^3$LO $NN$-only interaction evolved to the same $\lambda$ value. There, working only with a two-body potential, we were able to obtain results for $N_{\rm max}$ values as high as 17, but no substantial differences were found with respect to those obtained for $N_{\max}{=}13$.  Based on these considerations, we do not expect that an $N_{\rm max}=15$ calculation, which at this time is computationally out of reach, would significantly change the present  $N_{\rm max}=13$ picture for the phase shifts obtained with the SRG-evolved chiral $NN$+$3N$ interactions. Finally, we note that the NCSM ground-state energies for $\lambda=2.0\,\text{fm}^{-1}$ are well-converged with respect to $N_{\text{max}}$. The change between the $N_{\text{max}}=10$ and $12$ is below $0.2\%$. The same is true for $\lambda=1.88\,\text{fm}^{-1}$, where this change is below $0.1\%$, whereas for $\lambda=2.24\,\text{fm}^{-1}$ it is below $0.7\%$. 
 
\subsubsection{Dependence on $E_{3\rm{max}}$}
\begin{figure}[t]
\includegraphics*[width=1\columnwidth]{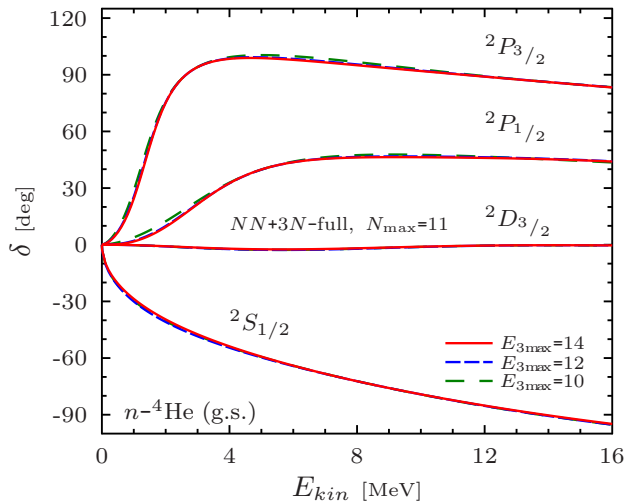}
\caption{
(color online) Dependence of the $n$-$^4$He phase shifts on $E_{3\rm max}$. Green long-dashed lines, blue dashed lines and red solid lines correspond to $E_{3\text{max}}= 10, 12$ and 14, respectively. The HO model space is truncated at $N_{\rm max}{=}11$ and all remaining parameters are as in Fig.~\ref{fig:Nmax-convergence}.} \label{fig:NA3max-convergence}
\end{figure}
In Fig.~\ref{fig:NA3max-convergence}, we study the variation of the $N_{\rm max}{=}11$ $n$-$^4$He(g.s.)~phase shifts with respect to the truncation of the $3N$ matrix elements $E_{3\rm{max}}$ as introduced in Sec.~\ref{sec:Hamiltonian&SRG}. This parameter is varied from $10$  to $14$ HO quanta, which is the largest set of $3N$ matrix elements currently available. Although at  $ N_{\rm max}{=}$11 a complete calculation would require $3N$ matrix elements with $E_{3\rm max}$ as high as 22 (see Sec.~\ref{sec:Hamiltonian&SRG}), the effects of this truncation are limited. The phase shifts at $E_{3\rm max}{=}12$  and  $E_{3\rm max}{=}14$ agree very well. A similar study has been repeated for different combinations of $\hbar \Omega, \lambda$, and $N_{\rm max}$. In general, we find that the $E_{3\text{max}}$ truncation of the three-nucleon interaction has less influence on the phase shifts than the parameters characterizing the NCSM/RGM model space. However, given that this limitation is not consistent with the adopted NCSM/RGM model-space, in any new application the dependence on $E_{3\rm max}$ should be analyzed with great attention. In the following, all results were obtained using the largest available $E_{3\rm max}{=}14$. 

\begin{table}[b]
\caption{{\em Ab initio} NCSM energies of the $^4$He first seven eigenstates (in MeV) at $N_{\rm max}{=}$12 (13 for negative parity) for different values of the SRG resolution scale computed with the $NN$+$3N$-full Hamiltonian at $\hbar\Omega=20\,\text{MeV}$. \label{table:NCSM-energies}}
\begin{ruledtabular}
\begin{tabular}{l|c|c|c|c|c|c|c}
$\lambda$ [fm$^{\texttt-1}$] &	g.s.		&		0$^+$0	&	0$^-$0	&	2$^-$0	&	2$^-$1   &   1$^-$1   &   1$^-$0\\
\hline
$2.24$       & -28.36 & -5.37 & -6.38 & -5.24 & -3.86 & -3.58 & -2.73 \\
$2.0$         & -28.44 & -5.62 & -6.51 & -5.39 & -4.03 & -3.77 & -2.95\\
$1.88$	  & -28.46 & -5.70 & -6.55 & -5.44 & -4.09 & -3.84 & -3.03\\
\hline
\hline
Expt.		    & -28.29 &  -8.08 & -7.28 & -6.45 & -4.99  & -4.65 & -4.04 
\end{tabular}
\end{ruledtabular}
\end{table}

\begin{figure}[t]
\includegraphics*[width=1\columnwidth]{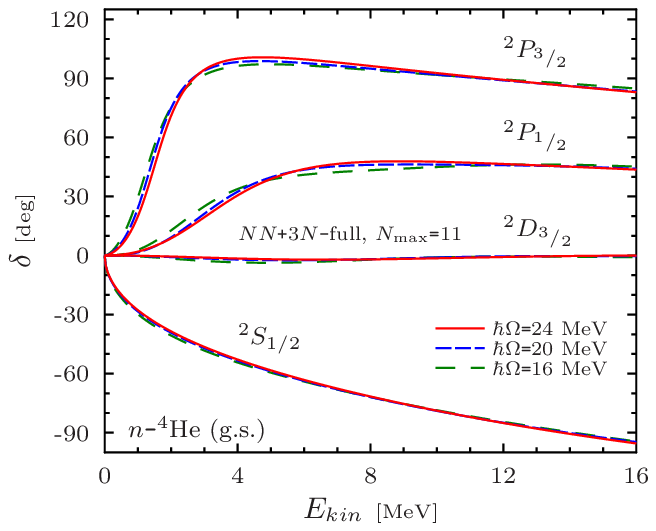}
\caption{(Color online) Dependence of the $n$-$^4$He phase shifts on the HO frequency. Green long-dashed lines, blue dashed lines and red solid lines correspond to $\hbar\Omega= 16, 20$ and 24, respectively. The model space is truncated at $N_{\rm max}=11$ and all remaining parameters are as in Fig.~\ref{fig:Nmax-convergence}.}\label{fig:omega-convergence}
\end{figure}

\subsubsection{Dependence on the HO frequency}

In Fig.~\ref{fig:omega-convergence}, we study the dependence of the $n$-$^{4}$He(g.s.) scattering phase shifts with respect to the HO frequency. We compare $\hbar\Omega{=}16$, $20$ and $24$ MeV results obtained with $\lambda {=}2.0~ \rm fm^{\texttt-1}$ in a $N_{\rm max}{=}11$ model space. We find essentially no dependence on the HO frequency when comparing the $\hbar\Omega=20$ and $24$ MeV, i.e., the phase shifts are in good agreement: the $^2 S_{1/2}$ and $^2 D_{3/2}$ phase shifts are on top of each other while the $^2 P_{1/2}$ and $^2 P_{3/2}$ phase shifts show very small deviations around the resonance positions. At the same time, we note that using the lower frequency of $\hbar\Omega=16\,\text{MeV}$ is problematic due to the finite size of the HO model space used during the SRG transformation. This has to be cured using a frequency conversion technique \cite{Roth2013}. For the remainder of this paper, we will stick to the $\hbar\Omega=20\,$MeV HO frequency. 

\begin{figure}[t]
\includegraphics*[width=1\columnwidth]{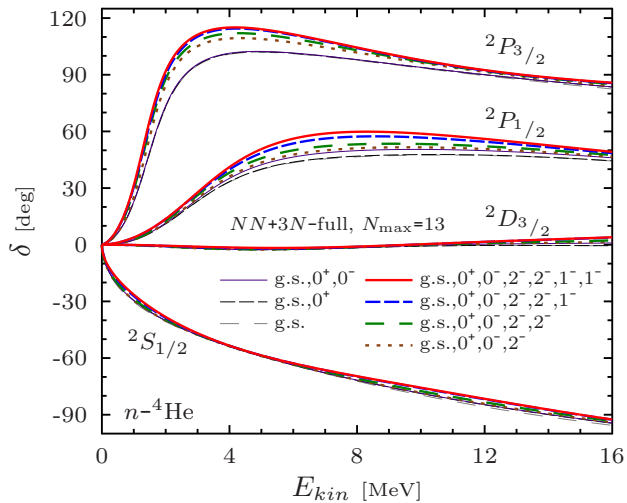}
\caption{
(color online) Dependence of the $n$-$^4$He phase shifts 
on the considered target eigenstates. Results with only the g.s.\ of $^4$He (thin gray long-dashed lines) are compared to those obtained by including in addition up to the 0$^+$0 (thin black dashed lines), 0$^-$0 (thin violet lines), 2$^-$0 (thick brown dotted lines), 2$^-$1 (thick green long-dashed lines), 1$^-$1 (thick blue dashed lines), and 1$^-$0 (thick red lines) excited states of \elem{He}{4}, respectively. The model space is truncated at $N_{\rm max}{=}13$. Other parameters are identical to those of Fig.~\ref{fig:Nmax-convergence}.}\label{fig:RGM-convergence}
\end{figure}

\begin{figure*}[t]
\includegraphics[width=1\textwidth]{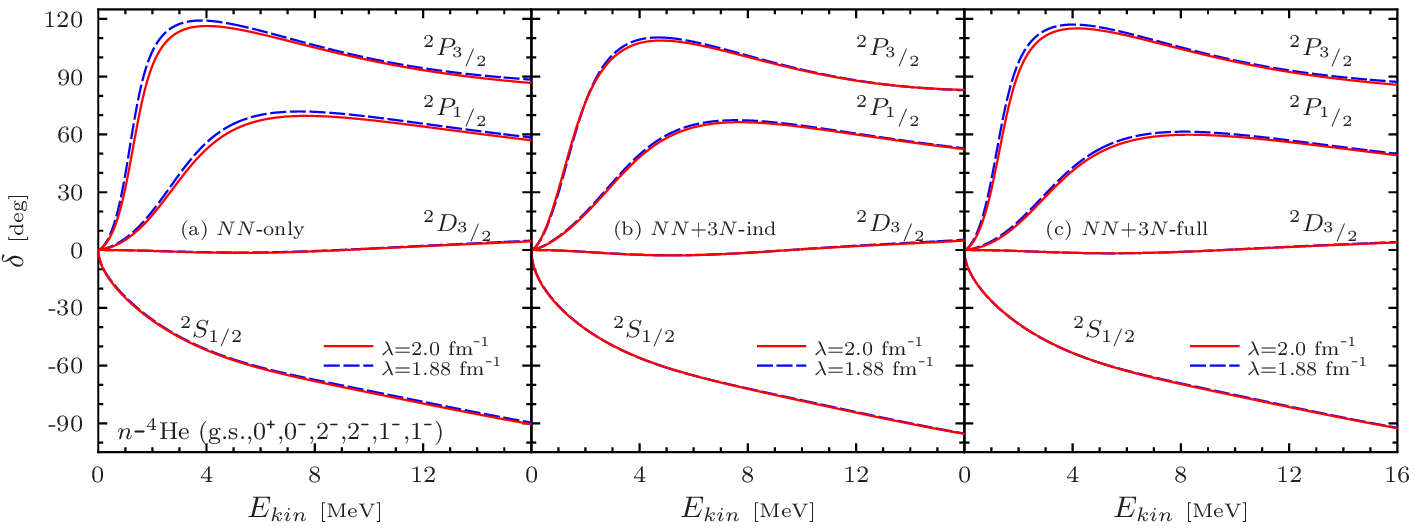}
\caption{(Color online) Comparison of the low-energy phase-shifts $^1S_{1/2}$, $^2P_{1/2}$, $^2P_{3/2}$ and $^2D_{3/2}$ of the $n$-$^{4}$He scattering for $\lambda {=}2.0~ \rm fm^{\texttt-1}$ (red solid lines) and $1.88~ \rm fm^{\texttt-1}$ (blue dashed lines) when accounting for the first six excited states of the $^4$He. Panel (a), (b) and (c) display the phase shifts obtained from the $NN$-only, $NN$+$3N$-induced and $NN$+$3N$-full Hamiltonians, respectively. The model space is truncated at $N_{\rm max}=13$. Remaining parameters are identical to Fig.~\ref{fig:Nmax-convergence}.}\label{fig:lambda-convergence}
\end{figure*}

\subsubsection{Dependence on excitations of the \elem{He}{4} target}

The many-body space spanned by the channel basis states of Eq.~(\ref{eq:ho-basis-n}) depends on the number of projectile and target eigenstates. In Fig.~\ref{fig:RGM-convergence}, we show the convergence pattern of the $n$-$^4$He phase shifts with respect to the inclusion of up to six low-lying excited states ($I_1^{\pi_1}T_1 = 0^+0,0^-0,2^-0,2^-1,1^-1$, and $1^-0$) of the $^4$He target, as obtained within the {\em ab initio} NCSM.  The energies of these states can be found in Table~\ref{table:NCSM-energies}. For this study, we employ our largest model space of $N_{\rm max}{=}13$. As can be seen in the figure, the target excitations are crucial in particular for the resonant phase shifts, where they lead to an enhancement. Specifically, the resonance of the $^{2}P_{3/2}$ wave is strongly influenced by the inclusion of the $2^-0$ state, while the $^{2}P_{1/2}$ phase shift is slowly enhanced near its resonance with the addition of the odd-parity states, among which the $1^-$ states have the strongest effect. On the other hand, the $^{2}S_{1/2}$ wave is Pauli blocked and mostly insensitive to these polarizations effects. The somewhat slow convergence displayed in Fig.~\ref{fig:RGM-convergence} is in agreement with that observed in Refs.~\cite{Quaglioni2008} and~\cite{Quaglioni2009}. In a recent work using a realistic two-body Hamiltonian~\cite{Baroni2013}, we have shown that a more efficient description of short and medium-range correlations, and hence a faster convergence, can be obtained within the NCSMC approach  \cite{Baroni2013} by augmenting the NCSM/RGM basis with NCSM $A$-body eigenstates of the compound (here $^5$He) system. Work is currently underway to complete the inclusion of $3N$-force contributions within (the couplings between $A$-nucleon eigenstates and binary-cluster basis states of) this approach. 

\subsubsection{Dependence on the SRG flow parameter}\label{sec:unitarity}

\begin{figure}[t]
\includegraphics[width=1\columnwidth]{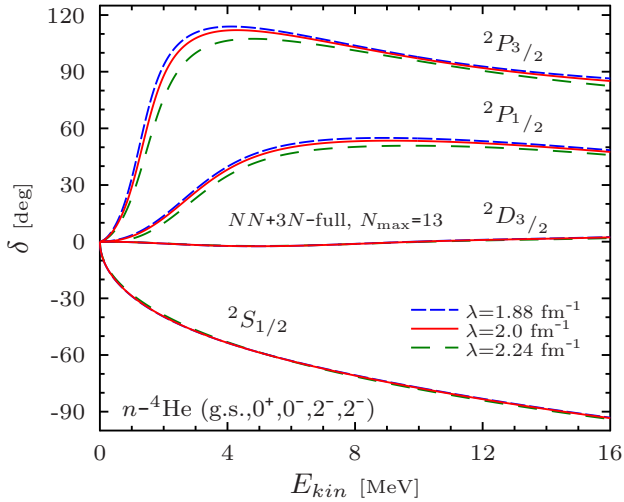}
\caption{
(Color online) Phase shifts of $n$-\elem{He}{4} scattering as in Fig.~\ref{fig:lambda-convergence}(c) including a third SRG parameter $\lambda=2.24\,\text{fm}^{-1}$ (green long-dashed lines), all computed with four excited states of the \elem{He}{4} target.}\label{fig:lambda-convergence-2}
\end{figure}

In Fig.~\ref{fig:lambda-convergence}, we compare the low-energy phase-shifts of the $n$-$^{4}$He scattering obtained for two different SRG flow parameters at $N_{\text{max}} =13$, including the first six excited states of the \elem{He}{4} target. In panels (a), (b) and (c) we compare $\lambda=1.88\,\text{fm}^{-1}$ to $\lambda=2.0\,\text{fm}^{-1}$ results for phase shifts obtained with the $NN$-only, $NN$+$3N$-induced and $NN$+$3N$-full Hamiltonian, respectively. For the $^2 P _{1/2}$ and $^2P_{3/2}$ phase shifts we observe a slightly larger flow-parameter dependence in the $NN$-only results compared to the $NN+3N$-induced case. Nevertheless, the $^2D_{3/2}$ and $^2S_{1/2}$ phase shifts agree very well for the two SRG parameters. Here, it is important to note that the convergence with respect to the model space size is relevant. Indeed we find less flow-parameter dependence in Fig.~\ref{fig:lambda-convergence} than in studies including only the g.s.\ of \elem{He}{4}.
This is also highlighted in Fig.~\ref{fig:lambda-convergence-2}, where the comparison for the most relevant $NN$+$3N$-full case is extended to a third value of $\lambda=2.24$ fm$^{-1}$, working within a smaller model space with four $^4$He excited states. Here, the $\lambda$ dependence is once again almost inexistent in the $^2S_{1/2}$ and $^2D_{3/2}$ phase shifts, where the convergence is fully achieved for all three choices of the SRG evolution parameter. However, larger differences can be observed in the resonances, particularly in the $^3P_{1/2}$ partial wave, where the lower rate of convergence for the somewhat harder $\lambda=2.24$ fm$^{-1}$ Hamiltonian is emphasized. Compared to the $NN$+$3N$-induced phase shifts, in the $^2P_{1/2}$ and $^2P_{3/2}$ partial waves we find a slightly larger $\lambda$-dependence. Altogether, the dependence on the SRG parameter in the $NN$+$3N$-induced and $NN$+$3N$-full cases is somewhat smaller than in the $NN$-only case. This hints towards a minor effect of higher-order SRG-induced many-body forces for the phase-shifts discussed here, consistent with the findings in NCSM calculations of the ground-state energies of light nuclei \cite{Jurgenson2009, Roth2011}. However, we stress that significant $\lambda$-dependence can also originate from non-convergence or inconsistent truncations even if no SRG induced many-body forces are present. In the NCSM/RGM the $E_{3\text{max}}$ parameter is not consistent with the model space truncation and could also lead to such an artificial SRG parameter dependence. Moreover, the SRG parameter dependence of the input target states, which is represented in Table~\ref{table:NCSM-energies} in terms of the $\lambda$-dependence of the eigenenergies, is of course translated into the NCSM/RGM kernels, too. We conclude, that the remaining SRG parameter dependence is a smaller effect than, e.g., the effects coming from the inclusion of more excited states of the \elem{He}{4} target as evident from Fig.~\ref{fig:RGM-convergence}. We stick to the resolution scale $\lambda=2.0\,\text{fm}^{-1}$ for our further investigations.

\subsubsection{Treatment of the isospin symmetry in the NCSM/RGM}\label{sec:isospin_average}

The initial chiral $3N$ interaction that we use in the present calculations is isospin invariant. The chiral $NN$ interaction, on the other hand, includes both the isospin and charge symmetry breaking, and, of course, we also add the proton-proton Coulomb interaction. For the three-nucleon SRG evolution,  we switch off the Coulomb interaction and use the isospin averaging of the $NN$ interaction developed in Ref.~\cite{Kamuntavicius1999}. The resulting SRG-induced $3N$ interaction is then isospin invariant and the expressions~(\ref{NNN-ex-express}) and (\ref{NNN-ex-express2}) are fully equivalent. The treatment of the (SRG-evolved) $NN$ interaction, however, differs depending on the approach taken to calculate the $NN$ interaction integration kernels. While the approach in the spirit of Eq.~(\ref{NNN-ex-express}) accounts fully for the isospin non-conservation of the $NN$ interaction, the approach that utilizes the isospin projection analogous to Eq.~(\ref{NNN-ex-express2}), used, e.g., in the original NCSM/RGM paper~\cite{Quaglioni2009}, relies on an isospin averaging of the $NN$ interaction. In particular, the $T{=}1$ two-nucleon $NN$ interaction matrix elements are calculated according to
\begin{equation}\label{NN_aver}
\langle \bar{V}_{NN}^{T=1} \rangle = c_{pn} \langle V_{pn}^{T=1} \rangle + c_{pp} \langle V_{pp}^{T=1} \rangle + c_{nn} \langle V_{nn}^{T=1} \rangle \;,
\end{equation}
with the weight coefficients given in terms of the number of the projectile protons and neutrons $Z_P, N_P$ and the target protons and neutrons $Z_T, N_T$ as
\begin{eqnarray}\label{weight_coef}
c_{pn} &=& \frac{\textstyle\frac{1}{2}(Z_P N_T + N_P Z_T)}{Z_P Z_T + N_P N_T +\textstyle\frac{1}{2}(Z_P N_T + N_P Z_T)} \;,\nonumber \\
c_{pp} &=& \frac{Z_P Z_T}{Z_P Z_T + N_P N_T +\textstyle\frac{1}{2}(Z_P N_T + N_P Z_T)} \;, \nonumber \\
c_{nn} &=& \frac{N_P N_T}{Z_P Z_T + N_P N_T +\textstyle\frac{1}{2}(Z_P N_T + N_P Z_T)} \; . 
\end{eqnarray}
The above expressions were derived by counting the number of $T{=}1$ nucleon pairs with one of the nucleons in the projectile and the other in the target, realizing that a proton-neutron pair has a 50\% probability to be in the $T{=}1$ state and a 50\% probability to be in the $T{=}0$ state. 

\begin{figure}[t]
\includegraphics[width=1\columnwidth]{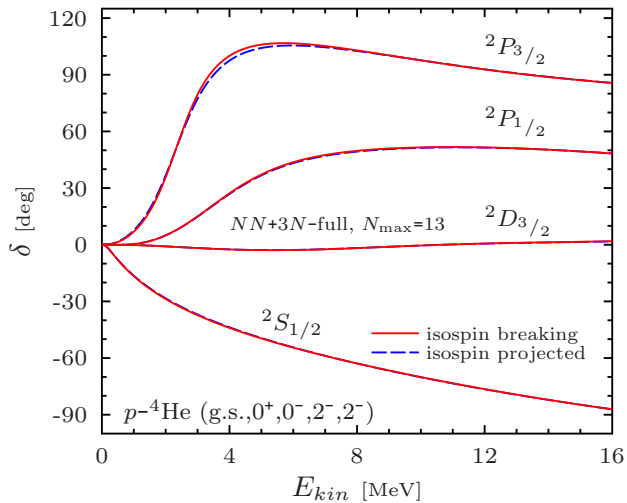}
\caption{
(Color online) Comparison of $p$-\elem{He}{4} phase shifts between the two approaches to compute the potential kernel, where the breaking of isospin symmetry is accounted for exactly as in Eq.~(\ref{NNN-ex-express}) (solid red lines) and with the approximation~(\ref{NN_aver}) and kernels alike Eq.~(\ref{NNN-ex-express3}) (blue dashed lines). The model space is truncated at $N_{\text{max}}=13$, the SRG parameter is $\lambda {=}2.0$ fm$^{\texttt-1}$ and $\hbar\Omega=20\,\text{MeV}$.}\label{fig:isospin}
\end{figure}

Since we now implemented also the $NN$ kernels in the spirit of Eq.~\eqref{NNN-ex-express}, i.e.~accounting the isospin-breaking effects, we tested the performance of the $NN$ interaction isospin averaging according to Eqs.~(\ref{NN_aver}) and (\ref{weight_coef}) in the $p$-$^4$He scattering, including up to the $2^-1$ excited state of the \elem{He}{4} target. The calculations with and without the isospin averaging are compared in Fig.~\ref{fig:isospin}. It is apparent that the approximations (\ref{NN_aver}) and (\ref{weight_coef}) are quite accurate, and, consequently, the isospin projected NCSM/RGM formalism is reliable as long as the isospin violation of the target wave functions is small, which is a typical case. Our calculations including six excited states of the \elem{He}{4} target, which are compared with experimental data in the following section, were obtained within the isospin-projected formalism.


\subsection{Discussion of results}\label{sec:results}

The extension of the NCSM/RGM to explicitly include $3N$ forces allows for the first investigation of SRG-evolved $NN$+$3N$ Hamiltonians in the energy continuum. This establishes access to a whole set of light-nucleus scattering observables as testing ground for chiral $NN$+$3N$ Hamiltonians. We start with a discussion about the $3N$ force effects on the $n$-\elem{He}{4} and $p$-\elem{He}{4} phase shifts and then compare those to experiment. We end this section with an investigation of cross sections and analyzing powers.

\subsubsection{Three-nucleon force effects on phase shifts}

\begin{figure}[t]
\includegraphics*[width=1\columnwidth]{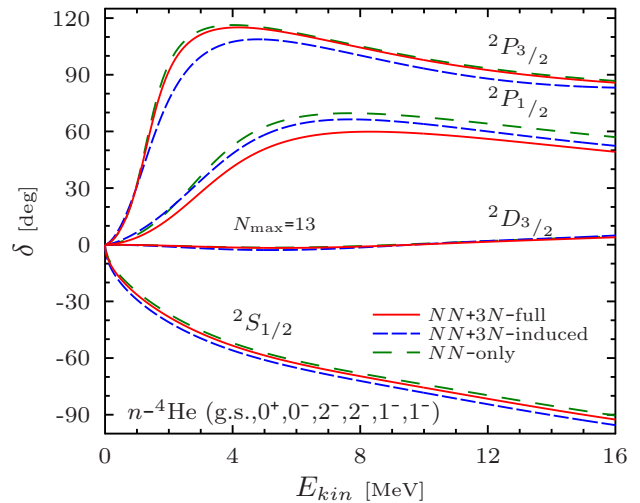}
\caption{
(Color online)  Comparison of the $n$-$^4$He phase shifts ($^1S_{1/2}$, $^2P_{1/2}$, $^2P_{3/2}$ and $^2D_{3/2}$ waves) for the three different interactions at $N_{\text{max}}=13$. The remaining parameters are identical to the ones of Fig.~\ref{fig:Nmax-convergence}.}\label{fig:NNN-comparison}
\end{figure}

In Fig.~\ref{fig:NNN-comparison}, we compare the $n$-\elem{He}{4} phase shifts for the $NN$-only (green long-dashed line), the $NN$+$3N$-induced (blue dashed line) and the $NN$+$3N$-full (red solid line) Hamiltonians. The NCSM/RGM model space includes the first six excited states of \elem{He}{4} and is truncated at $N_\text{max} = 13$, which is reasonably large as discussed in section~\ref{convergence}. The HO frequency is $\hbar\Omega=20\,\text{MeV}$ and the SRG flow-parameter $\lambda=2.0\,\text{fm}^{-1}$ is used.

Let us concentrate first on the $^2 S_{1/2}$ and $^2D_{3/2}$ partial waves. 
Comparing the $NN$-only phase shifts to those obtained with the $NN$+$3N$-induced Hamiltonian, we find they agree for the $^2D_{3/2}$ and only small changes occur for the $^2S_{1/2}$ partial wave. The same is true once the initial $3N$ interaction is included, i.e.~the $^2D_{3/2}$ and $^2S_{1/2}$ phase shifts are almost insensitive to the inclusion of $3N$ interactions. This is consistent with the observations of Ref.~\cite{Nollett2007}, where the AV18 $NN$ potential was complemented with the UIX or IL2 $3N$ force models. More influence of $3N$ interactions is present in the $^2P_{1/2}$ and $^2P_{3/2}$ phase shifts. The $NN$-only Hamiltonian overestimates the $^2P_{1/2}$ phase shift compared to $NN$+$3N$-induced Hamiltonian. The effect of the initial $3N$ interaction is a further decrease of this phase shift while the resonance gets slightly broadened. Concerning the $^2P_{3/2}$ partial wave, we find a quite large reduction of the phase shift in particular around the resonance position when including the induced $3N$ contributions. This hints at the significance of the SRG-induced $3N$ interactions and reveals an artificial enhancement of this phase shift using $NN$-only Hamiltonians. Including the initial $3N$ interaction increases this phase shift once again so that it almost coincides with the $NN$-only result, which is of course accidental. Overall we conclude that the inclusion of $3N$ interactions in scattering calculations is mandatory due to sizable changes in the $P$-wave phase shifts. In particular additional spin-orbit structures of the $3N$ interactions, which are well-known to be an essential part of the nuclear Hamiltonian \cite{Navratil2007a}, change the splitting between the $P$-wave resonances in $n$-\elem{He}{4} scattering. Finally, the net effect of the initial $3N$ force with respect to the results obtained with the initial $NN$ potential (unitarily equivalent to the $NN+3N$-induced results) is once again very similar to the trend observed in Ref. \cite{Nollett2007}, where the increased spin-orbit splitting between the resonances arises from an enhancement in the $3/2^-$ and quenching in the $1/2^-$ phase shifts, respectively.

\subsubsection{Comparison of phase shifts with experiment}

\begin{figure}[t]
\includegraphics*[width=1\columnwidth]{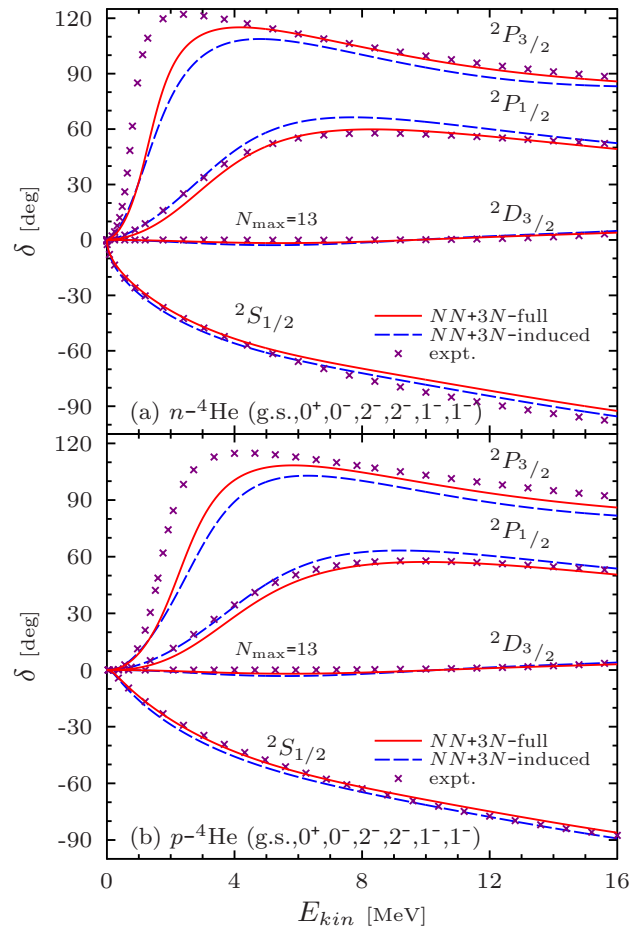}
\caption{
(color online) Comparison of the $n$-$^4$He (a) and $p$-\elem{He}{4} (b) phase-shifts ($^1S_{1/2}$, $^2P_{1/2}$, $^2P_{3/2}$ and $^2D_{3/2}$ waves) within the largest considered model space including the first six low-lying resonant states of the $^4$He (g.s., 0$^+$0, 0$^-$0, 2$^-$0, 2$^-$1, 1$^-$1, 1$^-$0) at $N_{\text{max}}=13$ to the experimental phase-shifts (purple crosses) obtained from an $R$-matrix analysis \cite{Hale}. Results for the $NN$+$3N$-full Hamiltonian are shown as red solid lines, those for the $NN+3N$-induced Hamiltonian as blue dashed lines. For remaining parameters see text or Fig.~\ref{fig:Nmax-convergence}.}\label{fig:expt-comparison-1}
\end{figure}

In the upper panel of Fig.~\ref{fig:expt-comparison-1} we compare our results for $n$-\elem{He}{4} low-energy phase shifts to experimental phase shifts obtained from an accurate $R$-matrix analysis of \elem{He}{5} data depicted as crosses \cite{Hale}. The phase shifts obtained with the $NN$+$3N$-full Hamiltonian are shown as solid red lines, while those from the $NN$+$3N$-induced Hamiltonian, which is unitarily equivalent to an initial $NN$ Hamiltonian, are given as dashed blue lines. Overall, we find a very good reproduction of the experimental phase shifts for the $^2S_{1/2}$, $^2P_{1/2}$ and $^2D_{3/2}$ partial waves. In particular for the $^2P_{1/2}$ phase shift the initial $3N$ interaction is responsible for this agreement. However, we also note that for the $^2S_{1/2}$ phase shift at large energies the initial $3N$ force leads to slightly larger deviations from experiment compared to the $NN$+$3N$-induced case. Concerning the $^2P_{3/2}$ partial wave, the initial $3N$ interaction brings the phase shift closer to experiment resulting in quite good agreement starting at about 4\,MeV c.m.\ energy. Moreover, this interaction leads to the expected increased spin-orbit splitting in the $P$ waves, as evident from the comparison to the splitting from the $NN$+$3N$-induced calculation. However, for energies around the resonance position the enhancement of the $^2P_{3/2}$ partial wave is too small, so that it misses the experimental resonance energy at 0.78\,MeV, related to the low-lying $\tfrac32^-\tfrac12$ state of $^5$He. While the need for an even richer spin-orbit structure of the $3N$ force cannot be ruled out \cite{Nollett2007}, one reason for this disagreement has to be searched in a still insufficient model space in our NCSM/RGM calculations for this partial wave. We have included the first six excited states of \elem{He}{4}, that is up to a maximum of 24\,MeV excitation energy (see Table \ref{table:NCSM-energies}). However, the $d$-$^3$H channel opens experimentally at $17.63$\,MeV, therefore, we would arrive at a more complete description of the scattering by including the coupling to this channel. This would require a generalization of the formalism developed in Ref.~\cite{Navratil2011} to include $3N$ forces. Another possibility to clarify this issue is to include the couplings to low-lying states of \elem{He}{5} in the framework of the NCSMC as recently proposed in Refs.~\cite{ Baroni2013} and~\cite{Baroni2013a}. Work in both these directions is currently underway.

In Fig.~\ref{fig:expt-comparison-1}(b) we present the comparison of the $p$-\elem{He}{4} phase shifts to those obtained from an $R$-matrix analysis of data~\cite{Hale}. The results are qualitatively very similar to the $n$-\elem{He}{4} case: The $^2S_{1/2}$ and $^2D_{3/2}$ partial waves are in good agreement with experimental phase shifts. This is also true for the $^2P_{1/2}$ phase shift at large energies including the initial $3N$ interaction. Instead, the initial $3N$ interaction  leads to deviations near the resonance located at 3.2\,MeV, where the calculation at the $NN$+$3N$-induced level is in good agreement with experiment. For the $^2P_{3/2}$ phase shift, the initial $3N$ interaction improves the overall agreement with experiment. However, we observe again quite large discrepancies at energies below 6\,MeV and the phase shift misses the experimental resonance of the \elem{Li}{5} system at 1.69\,MeV, though the spin-orbit splitting of the $^2P_{3/2}$ and $^2P_{1/2}$ phase shifts is increased by adding the initial $3N$ force. Again this deviation may be partly due to the limited model space and the missing $d$-\elem{He}{3} channel. 

Overall, the degree of agreement with experiment obtained for the nucleon-\elem{He}{4} phase shifts is very promising in view of future developments towards more refined reaction models and paves the way for \emph{ab-initio} applications and benchmarks of chiral $NN$+$3N$ Hamiltonians in nucleon scattering on light nuclei.

\subsubsection{Cross sections and analyzing powers}
\begin{figure}
\includegraphics*[width=1\columnwidth]{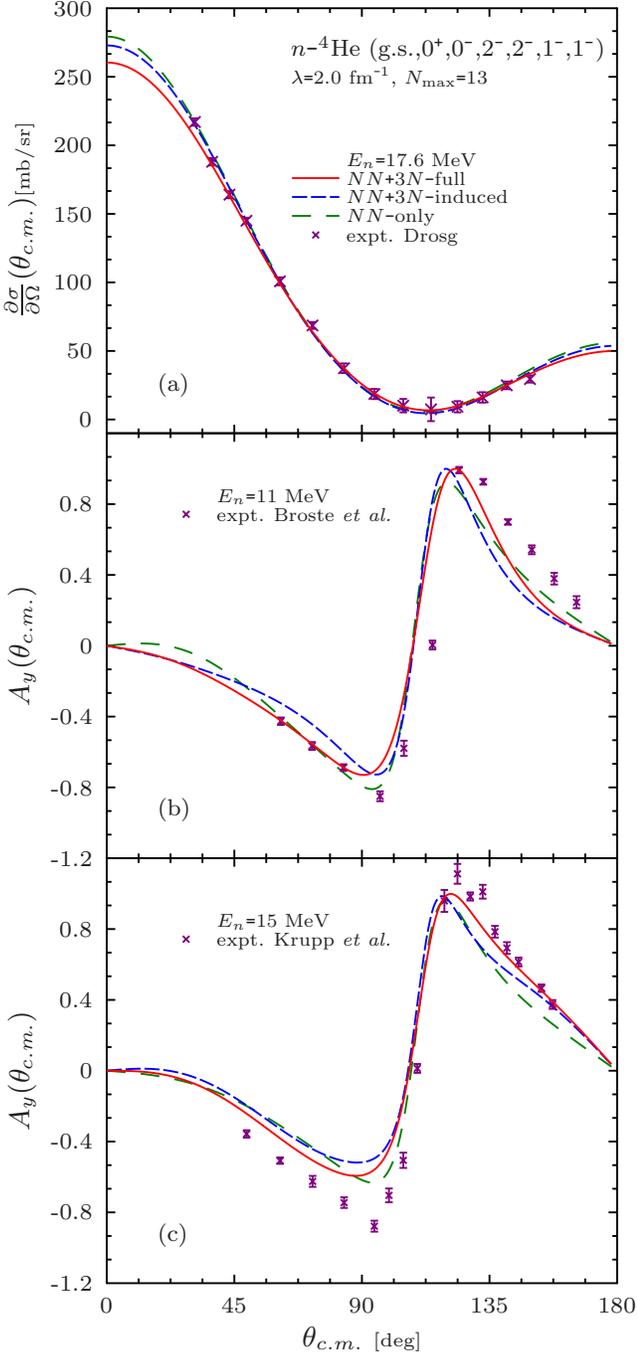}
\caption{(Color online) Comparison of $n$-\elem{He}{4} scattering differential cross-section at incident neutron energy $E_n=17.6$ MeV (a), and analyzing power at $E_n=11\,\text{MeV}$ (b) and $E_n=15\,\text{MeV}$ (c) to experiment~\cite{Drosg1978,Niiler1971,Broste1972,Krupp1984}. The different line styles correspond to the three different Hamiltonians. Remaining parameters are identical to the ones of Fig.~\ref{fig:expt-comparison-1}.}\label{fig:expt-comparison-2}
\end{figure}
\begin{figure*}
\includegraphics*[width=1\textwidth]{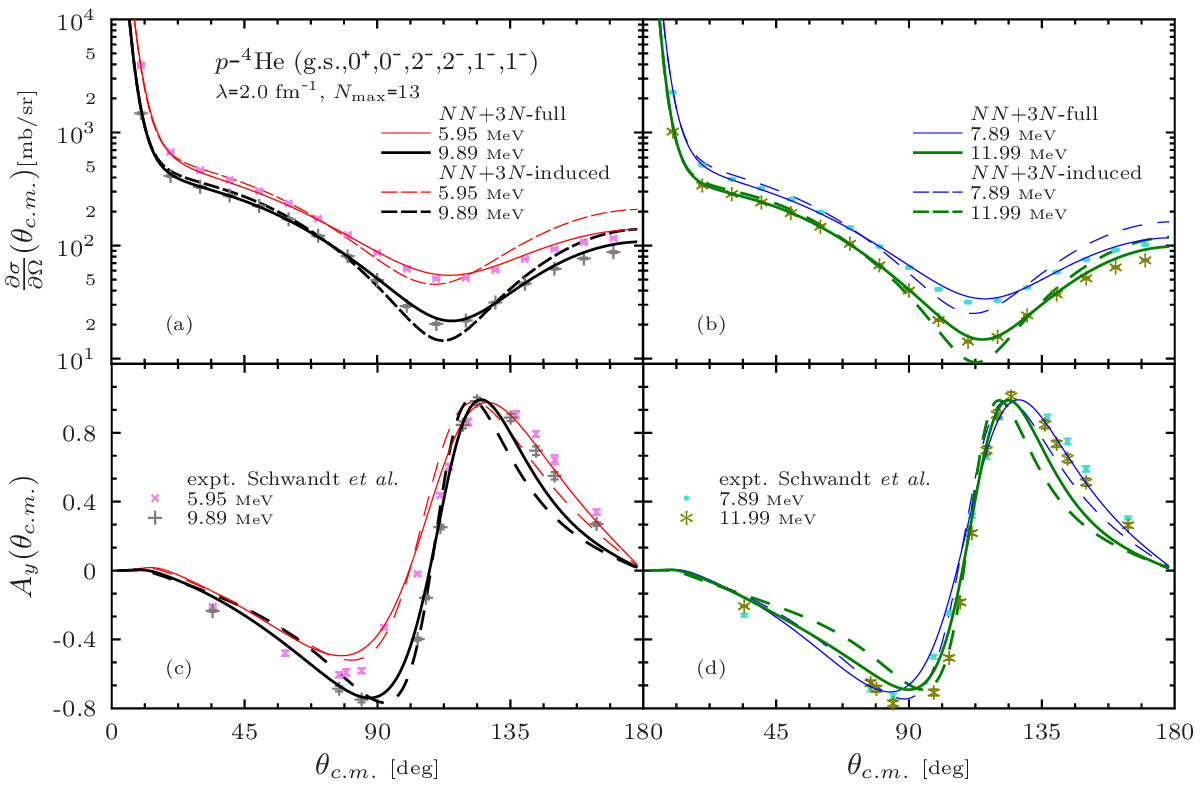}
\caption{
(Color online) Comparison of $p$-\elem{He}{4} scattering differential cross sections (upper panels) and analyzing powers (bottom panels) to experimental data~\cite{Schwandt1971}. Dashed lines represent results obtained with the $NN$+$3N$-induced Hamiltonian, solid lines those from the $NN$+$3N$-full Hamiltonian. The incident proton energies are $E_p = 5.95$ and $9.89$\,MeV in panels (a) and (c), and $E_p=7.89$ and $11.99$ MeV in panels (b) and (d), respectively. Remaining parameters are identical to Fig.~\ref{fig:expt-comparison-1}.}\label{fig:p-He4expt-comparison}
\end{figure*}
We now focus on a comparison of our theory to $n$-$^4$He and $p$-$^4$He elastic scattering observables. Using the chiral $NN$-only Hamiltonian,  in Ref.~\cite{Navratil2010} it was shown that NCSM/RGM calculations including only the g.s.\ and first excited state of the $^4$He target were able to provide a reasonable description of the angular differential cross section and analyzing power for these reactions at nucleon incident energies well beyond $10$ MeV. At those energies, as can be seen from Fig.~\ref{fig:RGM-convergence}, target polarization effects are weaker and, owing to the artificial  enhancement of the $^2P_{3/2}$ resonance with the $NN$-only Hamiltonian (see discussion of Sec. III.C.1),  one can reproduce the experimental scattering phase shifts already in such a smaller model space. As we will show in the following, in this work the inclusion of both SRG-induced and initial $3N$ forces, along with a larger model space containing $^4$He excited states up to $I_1^{\pi_1}T_1=1^-0$, extends the agreement with experiment down to about $4$ MeV c.m.\ energy. In Fig.\ref{fig:expt-comparison-2}(a), we compare the computed $NN$-only (green dashed lines), $NN+3N$-induced (blue long-dashed line), and $NN+3N$-full (red solid line) $^4$He$(n,n)^4$He differential cross section at an incident neutron energy of $17.6$ MeV to the experimental data of Drosg et al. \cite{Niiler1971, Drosg1978}. For all three Hamiltonians, we find good agreement at the angles where data are available. In particular, as one could expect based on the phase shifts of Fig.~\ref{fig:expt-comparison-1}, at this energy $3N$-force effects do not play a substantial role. Appreciable differences can be found only at large backward or forward angles, e.g., at angles below $30$ degrees the differential cross section decreases with the inclusion of both the induced and initial $3N$ interactions, resulting in a small underestimation at the first data point, but a slightly improved agreement in the last one. An overall similar behavior can be observed for the $p$-$^4$He cross sections at $E_p = 5.95$, $7.89$,  $9.89$, and $11.99$ MeV of Fig.~\ref{fig:p-He4expt-comparison}, panels (a) and (b). Here, the $NN+3N$-full curves are in excellent agreement with data at all angles, with the exception of large backward angles, where they slightly overestimate experiment. We reiterate that the $NN+3N$-induced Hamiltonian can be seen as unitarily equivalent to the initial chiral $NN$ one. In this sense, we can study the effects of the inclusion of the initial $3N$ interaction by comparing with the $NN$+$3N$-full curve. Different to the $n$-\elem{He}{4} differential cross sections, interestingly the $p$-\elem{He}{4} ones, for which data at intermediate incident energies are available, are more sensitive to the inclusion of the initial $3N$ interaction: for all considered energies the $NN$+$3N$-induced results show much larger deviations from experiment and in particular beyond 90 degrees the initial $3N$ force is responsible for the broad agreement.

The comparison with experiment is less straightforward for the analyzing power, as shown in Fig.~\ref{fig:expt-comparison-2}(b) and (c) for $n$-\elem{He}{4} scattering, where our results are plotted together with the $11$ and $15$ MeV data of Ref.~\cite{Broste1972} and Ref.~\cite{Krupp1984}, respectively. Compared to the $3N$-induced results, in this case the $NN$-only Hamiltonian (accidentally)  leads to better agreement with data around $90$ degrees while it deteriorates it at larger angles. 
Again we study the effects of the inclusion of the initial $3N$ interaction by comparing the $NN$+$3N$-induced to the $NN$+$3N$-full curve. This reveals that the initial $3N$ interaction improves the overall agreement with experiment  both around the dip at $90$ degrees and at larger angles, particularly in the $E_n=15$ MeV case. In general however, the discrepancies between theory and experiment are larger in this observable than they were in the differential cross section.
The reason for such more pronounced differences is twofold. First, the analyzing power is more sensitive to the spin-orbit components of the interaction \cite{Kievsky1999} as well as to possible shortcomings in the model space. 
For example, at the neutron laboratory energy of $11$ MeV the slight discrepancy with respect to experiment at the level of the phase shifts, particularly in the $^2S_{1/2}$ and $^2P_{3/2}$ partial waves, gets amplified in the $A_y$. Second, from $NN$-only calculations we know that at energies above $\sim11$ MeV one needs to include all partial waves up to at least $J=\tfrac{11}2$ to reach convergence in the partial wave expansion of the $A_y$. Here, however, we limit such an expansion to $J=\tfrac72$, owing to the fact that higher partial waves were found to be biased by the truncation of the $3N$ matrix elements, $E_{3{\rm max}}$. For $p$-$^4$He elastic scattering, where lower-energy data are available, the analyzing power at $E_p=7.89$\,MeV of Fig.~\ref{fig:p-He4expt-comparison}(d) computed with the $NN$+$3N$-full Hamiltonian is in good agreement with the measurements of Ref.~\cite{Schwandt1971} for all angles, and an equally good agreement around 90 degrees is found also at $E_p=9.89$\,MeV [see Fig.~\ref{fig:p-He4expt-comparison}(c)]. For $E_p = 5.95$\,MeV, which is closer to the resonance region, and $11.99\,\text{MeV}$ we see once again larger discrepancies in the dip region and starting at about $135$ degrees, respectively. The latter are in line with the previously observed overestimation of experiment in the corresponding differential cross section. Once again, except for the dip around 90 degrees at 5.95\,MeV neutron energy, the initial $3N$ force improves the agreement with experiment in comparison to our calculations using the $NN$+$3N$-induced Hamiltonian.

To conclude this section, we find that the adopted chiral $NN+3N$ Hamiltonian provides an overall good description of $n$-$^4$He and $p$-$^4$He elastic scattering away from the $^2P_{3/2}$ resonance, where the present model space is sufficient to reach convergence.


\section{Conclusions}
\label{sec:Conclusions}

We have presented the first application of $NN$+$3N$ interactions in {\em ab initio} scattering calculations within the NCSM/RGM approach for nucleon-nucleus collisions. We have generalized the formalism to explicitly include $3N$ interactions, which generated new contributions to the potential kernel. One of these contributions involves three-body density matrices, posing a formidable computational task. We have developed two efficient ways to deal with this complication and have used them to verify our results.

In this first application, we have analyzed nucleon-\elem{He}{4} scattering using consistently SRG-evolved chiral $NN$+$3N$ interactions, by performing a comprehensive sensitivity study of all relevant parameters in the NCSM/RGM approach. We found fairly good convergence with respect to the model-space size, particularly at intermediate energies, and only a minor dependence on the SRG flow parameter. 

Finally, we have compared our results to selected experimental data for phase shifts, angular differential cross sections and polarization observables obtained from our most complete calculation including the first six excited states of the \elem{He}{4} target in the NCSM/RGM model space. We found significant contributions of both the SRG-induced $3N$ and the initial $3N$ interactions in particular for the $P$-wave phase shifts. Overall, we have reached very good agreement with phase shifts obtained from accurate $R$-matrix analysis of the $A=5$ data. The spin-orbit splitting between the $P$-wave phase shifts is enhanced by the initial $3N$ interaction, but some discrepancies with experiment still remain for the $^2P_{3/2}$ phase shift in the energy region near the resonance.
Although the need for spin-orbit structures beyond those present in the chiral N$^2$LO $3N$ force cannot be ruled out, this is more likely due to a still insufficient model space in terms of excited states of the $^4$He target and will be addressed in the future by using the present NCSM/RGM kernels within the recently proposed NCSMC approach and, most optimally, by also including $d$-$^3$H ($^3$He) channels explicitly. For differential cross sections, we have found remarkably good agreement with experiment even over a wide range of incident nucleon energies, while in the case of the analyzing power we have encountered larger deviations at certain angles and energies.

Altogether, this work paves the way for {\em ab initio} benchmarks and applications of $NN$+$3N$ Hamiltonians in the context of light-nucleus reaction observables. Future work with the nucleon-nucleus formalism will address the role of the $3N$ force in four-nucleon reaction observables as well as scattering including heavier targets which is possible due to one of the newly developed kernel implementations presented here. The present NCSM/RGM kernels will be used within the NCSMC with explicit $3N$ interactions to revisit reactions of particular astrophysical interest. Work to include the $3N$ force within the NCSMC deuteron-nucleus formalism is also underway.


\begin{acknowledgments}
Prepared in part by LLNL under Contract DE-AC52-07NA27344. We acknowledge support from the U. S. DOE/SC/NP (Work Proposal No. SCW1158),  from the Deutsche Forschungsgemeinschaft through contract SFB 634, from the Helmholtz International Center for FAIR within the framework of the LOEWE program launched by the State of Hesse, from the BMBF through contract 06DA7074I and from the NSERC Grant No. 401945-2011. TRIUMF receives funding via a contribution through the Canadian National Research Council. Computing support for this work came from the LLNL institutional Computing Grand Challenge program, the J\"ulich Supercomputing Center, the LOEWE-CSC Frankfurt, and the National Energy Research Scientific Computing Center supported by the Office of Science of the U.S. Department of Energy under Contract No. DE-AC02-05CHH11231.
\end{acknowledgments}

\appendix


\section{$3N$ contribution to the direct kernel}\label{sec:appendix-direct}

In this appendix, we give the expressions of the direct potential kernel of Eq.~(\ref{pot-NNN-direct}) in the new basis defined by Eq.~(\ref{eq:new-basis-formula}) of the SD channel states. We start with the expression in the uncoupled scheme
\begin{widetext}
\begin{align}
\leftsub{\rm SD}{\big\langle}& \Phi^{J^\pi T}_{\kappa^\prime}\big|  V^{3N}_{A-2\,A-1\,A} \big(1-P_{A-1,A} - P_{A-2,A} \big) | \Phi^{J^\pi T}_{\kappa}\rangle_{\rm SD} \notag =\frac{1}{2(A-1)(A-2)} \sum_{M_1 m_j} \sum_{M_{T_1} m_t} \sum_{M^{\prime}_1 m^{\prime}_j} \sum_{M^{\prime}_{T_1} m^{\prime}_t}\\
 \times&\clebsch{I_1}{j}{J}{M_{1}}{m_j}{M_J} \clebsch{T_1}{\tfrac{1}{2}}{T}{M_{T_1}}{m_t}{M_T} \clebsch{I^{\prime}_1}{j^{\prime}}{J}{M^{\prime}_{1}}{m^{\prime}_j}{M^{\prime}_J} \clebsch{T^{\prime}_1}{\tfrac{1}{2}}{T}{M^{\prime}_{T_1}}{m^{\prime}_t}{M^{\prime}_T}  \notag \\
\times&\sum_{abde} 
 \leftsub{\rm SD}{\big\langle} A{-}1\, \alpha_1^{\prime} I_1'^{\pi_1^{\prime}} M_1^{\prime} T_1^{\prime} M_{T_1}^{\prime}\big| a^{\dagger}_{a} a^{\dagger}_{b}  a_{d} a_{e} \ket{A{-}1\alpha_{1} I_1^{\pi_1} M_1 T_1 M_{T_1}}_{\rm SD}   \;
\bra{ba\; n^{\prime}\ell^{\prime}j^{\prime}m_{j}^{\prime}\tfrac12 m_{t}^{\prime}} V^{3N}
 \ket{de\;n\ell j m_j \tfrac12 m_t}
\end{align}
\end{widetext}
using the same notations as in Eq.~\eqref{NNN-ex-express}.

Alternatively, we can switch to the scheme using coupled densities in the same way as discussed before [Eq.~\eqref{NNN-ex-express3}]. After the application of the Wigner-Eckart theorem the direct potential kernel of Eq.~(\ref{pot-NNN-direct}) simplifies to
\begin{widetext}
\begin{align}
\leftsub{\rm SD}{\langle} & \Phi^{J^\pi T}_{\kappa^\prime} | V^{3N}_{A-2,A-1,A}\big(1-P_{A-1,A} - P_{A-2,A} \big) | \Phi^{J^\pi T}_{\kappa}\rangle_{\rm SD} = \frac{1}{2(A{-}1)(A{-}2)} \sum_{{\bar a} {\bar b} {\bar d} {\bar e}} \sum_{J_{ab} T_{ab}} \sum_{J_{de} T_{de}} \sum_{J_0 T_0} \sum_{K \tau}\nonumber\\
\times &  ~ {\hat \tau} {\hat K} {\hat J_0}^2{\hat T_0}^2 ~ (-1)^{J_{de}+J_0+2j'+J+I_1+T_{de}+T_0+T+T_1}   \langle {\bar a}, {\bar b} ; J_{ab} T_{ab}, n'\ell' j' \tfrac12 ; J_0 T_0 | V^{3N}  |  {\bar d}, {\bar e} ; J_{de} T_{de},n\ell j \tfrac12 ; J_0 T_0 \rangle\nonumber\\ 
\times& ~ \left\{
\begin{array}{ccc}
I_1 & K & I_1' \\
j' & J & j
\end{array}
\right\} \left\{
\begin{array}{ccc}
J_{ab} & J_{de} & K \\
j & j' & J_0 
\end{array} 
\right\}\left\{
\begin{array}{ccc}
T_1 &\tau&T_1' \\
\tfrac12 & T & \tfrac12
\end{array}
\right\} \left\{
\begin{array}{ccc}
T_{ab}&T_{de} &\tau \\
\tfrac12 & \tfrac12 &T_0
\end{array} \right\} \nonumber\\
\times& ~\leftsub{\rm SD}{\langle} A{-}1\alpha_{1}' I_1^{\prime \, \pi_1'}T_1'  |||  \Big( \big( a^{\dag}_{{\bar a}}a^{\dag}_{{\bar b}} \big)^{J_{ab} T_{ab}} \big( {\tilde a}_{{\bar d}} {\tilde a}_{{\bar e}} \big)^{J_{de}T_{de}} \Big)^{K \tau} ||| A{-}1\alpha_{1} I_1^{\pi_1} T_1 \rangle_{\rm SD} \; .
\label{eq:V3N2BDcoupled}
\end{align}
\end{widetext}
As in Eq.~(\ref{NNN-ex-express3}) we have explicitly performed the summations over angular momentum and isospin projections using the Wigner-Eckart theorem and the same notations for HO orbitals and reduced density matrix elements.

\bibliographystyle{apsrev4-1}
\bibliography{biblio}

\end{document}